 \documentclass[final,3p,times,authoryear]{elsarticle}

\usepackage{amssymb}
\usepackage{amsmath}

\usepackage{amsthm}

\usepackage{multirow}
\usepackage{textcomp}
\usepackage{subcaption} 
\usepackage{siunitx}
\usepackage{epstopdf}
\usepackage[table,xcdraw]{xcolor}
\usepackage{url} 
\usepackage{hyperref}
\journal{Nuclear Physics B}

\usepackage{amssymb}  
\usepackage{booktabs} 
\usepackage{makecell} 

\begin{document}

\begin{frontmatter}



\title{AnalogMaster: Large Language Model-based Automated Analog IC Design Framework from Image to Layout} 





\author[a,1]{Xian Rong Qin}
\ead{qinxianrong@whut.edu.cn}

\author[a,b,1]{Yong Zhang}
\ead{yongz@whut.edu.cn} 

\author[a]{Ying Hu}
\ead{hu_ying@whut.edu.cn}

\author[a]{Tao Su}
\ead{sutao0606@whut.edu.cn}

\author[a,b]{Bo-Wen Jia\corref{cor1}}
\ead{jiabowen@whut.edu.cn}

\author[a,b]{Ning Xu\corref{cor1}}
\ead{xuning@whut.edu.cn}

\cortext[cor1]{Corresponding author}

\fntext[1]{These authors have contributed equally to this work and are co-first authors.}

\affiliation[a]{organization={School of Information Engineering, Wuhan University of Technology},
                addressline={},
                postcode={430070},
                city={Wuhan},
                country={China}}

\affiliation[b]{organization={State Key Laboratory of Integrated Circuits and Systems, Fudan University},
                addressline={},
                postcode={200433},
                city={Shanghai},
                country={China}}


\begin{abstract}
Design automation has the potential to substantially improve the efficiency of analog integrated circuit (IC) design. However, existing algorithms and tools typically focus on individual stages, such as device sizing, placement, or routing, and still require significant manual intervention to complete the full design flow. While large language models (LLMs) have recently demonstrated remarkable success in automating digital IC design workflows, these advances cannot be directly transferred to analog IC design. Key challenges include strongly coupled performance metrics, the predominance of unstructured circuit schematic images, and the fact that most prior approaches address only isolated stages of the analog design process, limiting their ability to capture end-to-end performance impact. To address these challenges, we propose AnalogMaster, an extensible, LLM-based framework that enables end-to-end automation of analog IC design through a unified pipeline spanning circuit image–to–netlist generation, parameter optimization, placement, and routing. AnalogMaster integrates a joint reasoning mechanism that leverages in-context learning and intent reasoning to achieve accurate and robust image-to-netlist conversion. A parameter search agent integrating self-enhanced prompt engineering and context truncation is developed for effective device sizing and downstream physical design.  Experimental evaluations on 15 representative circuits with varying levels of complexity demonstrate strong and consistent performance across multiple models. In particular, GPT-5 achieves success rates of 92.9\% and 99.9\% on Pass@1 and Pass@5, respectively. These results validate the effectiveness and robustness of the proposed framework and establish a practical paradigm for applying LLMs to full-stack analog IC design automation.
\end{abstract} 
\label{sec:Abstract}

\begin{graphicalabstract}
\end{graphicalabstract}

\begin{highlights}
\item Research highlight 1:
Proposed AnalogMaster, the first LLM-based framework for end-to-end analog IC design automation.

\item Research highlight 2:
Introduced a joint reasoning mechanism to enhance netlist extraction from complex circuit images.

\item  Resarch highlight 3:
Designed a parameter search agent to auto-compress device parameter space for downstream physical design.

\item  Rearch highlight 4:
Demonstrated state-of-the-art performance across multiple MLLMs on a 15-circuit benchmark.

\end{highlights}


\end{frontmatter}



\section{Introduction}

The rapid advancement of large language models (LLMs) has enabled significant breakthroughs in natural language processing and general-purpose code generation\citep{JIN20253961}. Inspired by these successes, recent research has begun to explore the application of LLMs to Electronic Design Automation (EDA). Benefiting from the high-level abstraction in hardware description languages (HDLs) and the availability of large-scale, high-quality open-source datasets, LLM-based approaches have made notable progress in digital circuit design. Representative efforts include: (1) HDL code generation and benchmark evaluation \citep{liu2023verilogeval,thakur2023autochip,lu2024rtllm,thakur2024verigen,liu2024rtlcoder}; and (2) automated bug detection, repair, and testbench/script synthesis \citep{qiu2024autobench,bhandari2024llm-aided-testbench,liu2023chipnemo}. In contrast, automated analog integrated circuit(IC) design presents fundamentally different challenges. Key performance metrics, such as gain, bandwidth, power consumption, and noise, exhibit strong nonlinear dependencies and complex cross-coupling effects. These intricate interdependencies propagate throughout the entire design flow—spanning netlist generation, device sizing, placement, and routing—rendering each stage highly sensitive to decisions made in others. Consequently, even a single circuit topology must be meticulously adapted to meet diverse application-specific specifications, requiring deep domain knowledge and holistic co-optimization across all stages. Isolated tuning of any individual step is insufficient; moreover, as circuit complexity increases, the combinatorial explosion of cross-stage constraints renders manual, stage-by-stage optimization extremely time-consuming and labor-intensive. This reliance on human expertise not only limits scalability but also underscores the formidable difficulty—and pressing necessity—of achieving a fully automated, end-to-end analog IC design methodology.

Netlists in analog IC design are typically constructed interactively by engineers using commercial platforms such as Virtuoso. Recently, several studies have explored netlist generation using learning-based approaches. AnalogCoder \citep{lai2025analogcoder} proposes a training-free LLM agent that translates natural language specifications into netlists via intermediate Python code, while SpicePilot \citep{vungarala2024spicepilot} employs domain-informed prompting to iteratively guide PySpice/SPICE netlist synthesis. Compared with digital IC design, analog IC design lacks large-scale, well-structured datasets; instead, a substantial portion of analog design knowledge is embedded in schematic diagrams. Consequently, directly extracting netlists from circuit schematics is essential for advancing the application of LLMs in analog IC EDA. Moreover, representing circuit structures through general-purpose programming languages incurs substantial token overhead and higher syntactic error rates compared with native SPICE formats, which degrades generation accuracy and complicates downstream design stages.  Even with accurate netlist extraction, achieving high-performance analog ICs requires precise device sizing—an area where automation remains largely decoupled from topology synthesis. In the domain of device sizing, \citep{ROJEC201948} proposed an evolutionary algorithm-based approach that simultaneously optimizes both the topology and component parameters of analog circuits through a novel graph-based representation and hybrid parameter optimization. MACE\citep{lyu2018batch} proposes a batch Bayesian optimization framework that parallelizes optimization by integrating multiple acquisition functions into a multi-objective optimization setting. \citep{somayaji2021prioritized}introduces a prioritized reinforcement learning (RL) framework that combines non-uniform prioritized experience replay, a two-stage critic network embedding design knowledge, and trajectory-guided local exploration. AmpAgent \citep{liu2024ampagent} further adopts a multi-agent collaborative architecture that integrates literature mining, mathematical reasoning, and hybrid optimization (ABC/TuRBO) to design multi-stage amplifiers. Although these methods demonstrate impressive results in device sizing, they rely on manually specified circuit topologies or initial netlists and do not extend to placement and routing based on optimized parameters. As a result, their impact on overall circuit performance and physical feasibility remains difficult to assess.  For analog IC placement,  \citep{ma2010simultaneous}presents a simulated annealing (SA) based framework that incorporates sequence pairs and center-based corner block lists to handle symmetry, common-centroid, and general placement constraints.   \citep{xu2017hierarchical} proposes a hierarchical analytical placement technique for high-performance analog ICs, achieving significant reductions in both critical net wirelength and layout area compared with conventional approaches and manual designs.  Nevertheless, these methods require a human-provided netlist as input and do not proceed to post-placement routing, leaving routing feasibility and post-layout circuit performance unverified. In the routing domain, \citep{zhang2025reinforcement} introduces a design-rule-compliant bidirectional A* algorithm combined with an attention-augmented RL framework for track assignment, improving routing efficiency. \citep{zhang2025pin} proposes a pin-access-planning-driven matching routing methodology for analog ICs, enhancing both routing symmetry and layout performance. \citep{xu2025paroute2} presents PARoute2, a performance-driven analog routing paradigm that leverages machine learning to address the limitations of traditional methods. By integrating a 3DGNN-based performance predictor and a 3DGRU-based routing guidance generator, PARoute2 generates nonuniform routing guidance to optimize post-layout performance metrics. However, these techniques operate under the assumption of a given netlist and pre-computed placement, and thus do not constitute an end-to-end solution from specification to physical implementation.

In summary, existing studies predominantly decompose analog IC design into isolated subtasks, such as netlist generation, sizing, placement, and routing, then optimize each component independently. Although these approaches make meaningful technical progress, they neglect the strong interdependencies across the analog design flow, thereby obscuring how local design decisions propagate to global performance and physical feasibility. As summarized in  Table\ref{tab:method_comparison}, few previous works have achieved end-to-end automation spanning circuit image interpretation, parameterized netlist synthesis, performance optimization, and placement and routing. This fragmentation creates a critical gap that limits both the practical applicability of current methods and the systematic evaluation of LLM capabilities in analog EDA. To address this gap and fully exploit the potential of LLMs across the entire analog design lifecycle, we propose \textbf{AnalogMaster}, an extensible LLM-based framework for full-stack analog IC automation.

\begin{table}[htbp]
\centering
\caption{Comparison of Recent LLM-based Methods for Analog IC Design}
\label{tab:method_comparison}
\small
\begin{tabular}{l|ccccccc}
\toprule
\textbf{Method} & \textbf{\makecell{Image-Netlist\\Generation\textsuperscript{a}}} & \textbf{Sizing} & \textbf{Placement} & \textbf{Routing} & \textbf{\makecell{Fully\\Automated\textsuperscript{b}}} & \textbf{Benchmarks\textsuperscript{c}} \\
\midrule
AnalogCoder\textsuperscript{1}     &  $ \times $  &  $ \times $  &  $ \times $  &  $ \times $  &  $ \checkmark $  &  $ \checkmark $    \\
SPICEPilot\textsuperscript{2}      &  $ \times $  &  $ \times $  &  $ \times $  &  $ \times $  &  $ \times $      &  $ \times $        \\
LaMAGIC\textsuperscript{3}         &  $ \times $  &  $ \times $  &  $ \times $  &  $ \times $  &  $ \checkmark $  &  $ \times $        \\
LLM-USO\textsuperscript{4}         &  $ \times $  &  $ \checkmark $  &  $ \times $  &  $ \times $  &  $ \checkmark $  &  $ \times $        \\
AnalogXpert\textsuperscript{5}     &  $ \times $  &  $ \times $  &  $ \times $  &  $ \times $  &  $ \checkmark $  &  $ \checkmark $    \\
MasaCHAI\textsuperscript{6}        &  $ \checkmark $  &  $ \times $  &  $ \times $  &  $ \times $  &  $ \checkmark $  &  $ \checkmark $    \\
LADAC\textsuperscript{7}           &  $ \times $  &  $ \checkmark $  &  $ \times $  &  $ \times $  &  $ \times $      &  $ \times $          \\
AmpAgent\textsuperscript{8}        &  $ \times $  &  $ \checkmark $  &  $ \times $  &  $ \times $  &  $ \times $      &  $ \times $         \\
LayoutCopilot\textsuperscript{9}   &  $ \times $  &  $ \times $  &  $ \checkmark $  &  $ \times $  &  $ \times $      &  $ \times $          \\
\addlinespace[1pt]
\textbf{AnalogMaster}              &  $ \checkmark $  &  $ \checkmark $  &  $ \checkmark $  &  $ \checkmark $  &  $ \checkmark $  &  $ \checkmark $  &   \\
\bottomrule
\end{tabular}

\par\vspace{2pt}
\footnotesize
\textsuperscript{a} Convert the circuit image to the netlist. 
\textsuperscript{b} Without human intervention. 
\textsuperscript{c} Provided benchmarks.
\noindent\footnotesize
\textsuperscript{1}\cite{lai2025analogcoder}, 
\textsuperscript{2}\cite{vungarala2024spicepilot}, 
\textsuperscript{3}\cite{chang2024lamagic}, 
\textsuperscript{4}\cite{somayaji2025llm-uso}, 
\textsuperscript{5}\cite{zhang2025analogxpert}, 
\textsuperscript{6}\cite{masalaCHAI-2025}, 
\textsuperscript{7}\cite{liu2024ladac}, 
\textsuperscript{8}\cite{liu2024ampagent}, 
\textsuperscript{9}\cite{liu2025layoutcopilot}.
\end{table}

Our main contributions are:

1. We propose AnalogMaster, an extensible LLM-based framework that enables end-to-end analog IC design automation, including circuit images-to-netlist conversion, sizing, placement, and routing. This work establishes a new paradigm for integrating LLM into the complete IC design flow.

2. We address the high error rate in directly converting circuit images to SPICE netlists by introducing a joint reasoning mechanism, which substantially improves the model’s netlist extraction accuracy from complex circuit schematics by synergistically integrating multi-dimensional reasoning cues derived from circuit images.

3. We develop a parameter search agent that effectively compresses the device parameter search space for visually extracted netlists, addressing the limitation that topology-only netlists can't be directly used for placement and routing. Compared with traditional methods that rely on manually defined parameter ranges, the proposed approach is more general and adaptive. 

4. Experimental results demonstrate that the AnalogMaster framework consistently achieves outstanding performance across multiple MLLMs on a benchmark comprising 15 representative circuit diagrams with varying difficulty levels.  In particular, GPT-5 achieves success rates of 92.9\% on Pass@1 and 99.9\% on Pass@5, validating the effectiveness and robustness of the proposed framework.  
\label{sec:introduction}

\section{Background}

\subsection{Multimodal Large Language Models \& Agent}

Multimodal Large Language Models (MLLMs) are pretrained on massive generic corpora of images and text, enabling strong cross-modal generalization through aligned token representations: given an image $\mathbf{x}_{\text{img}}$ and a text sequence $\mathbf{x}_{\text{text}}$, they construct a joint context $\mathbf{Z} = [\mathbf{E}_{\text{text}}, \mathbf{E}_{\text{img}}]$ for autoregressive decoding. This architecture supports in-context learning\citep{min2022metaicl}, allowing the model to infer task intent from few-shot demonstrations without parameter updates. However, when applied to analog IC design---a domain with scarce, expert-dependent annotations and highly structured circuit images---MLLMs face severe limitations. Their pretraining data lack the domain-specific semantics of circuit topology, making direct end-to-end netlist generation from raw diagrams unreliable. Recent studies \citep{masalaCHAI-2025}\citep{liu2024ampagent} confirm that even state-of-the-art MLLMs fail to consistently recover correct netlists.  

MLLMs are fundamentally stateless, probabilistic systems: each inference step processes input and generates output as independent high-dimensional vector transformations, without intrinsic memory or sustained reasoning capabilities. However, many analog IC design tasks—such as circuit netlist synthesis, sizing, placement, and routing—demand extended cognitive functions like long-term memory, autonomous planning, environmental awareness, and dynamic decision-making. Successful execution requires iterative strategy formulation, anomaly detection, result validation, and refinement until the final objective is achieved. To bridge this gap, recent works\citep{liu2024ampagent}\citep{wu2024chateda}\citep{somayaji2025llm-uso} have augmented agents by integrating LLMs with external memory modules, tool invocation interfaces, state-tracking mechanisms, and specialized prompt engineering strategies. These enhancements significantly extend the practical applicability of LLMs in EDA contexts. Additionally, multi-agent collaborative frameworks\citep{liu2024ampagent} have been proposed to enable role specialization and distributed reasoning, thus facilitating the automation of more sophisticated and higher-order EDA functions.

However, both MLLMs and agents remain constrained by finite context windows: excessive input truncates critical information, while insufficient context impairs coherence. In a complex analog circuit image, these constraints amplify the risk of hallucination---generating syntactically valid but fictional or irrelevant netlists that do not correspond to the input image, leading to functionally incorrect or non-existent circuits.

\subsection{Automated process for analog IC design}

The analog IC design process can be divided into the following five stages. 

\textbf{1) Circuit Design}. Traditional design mandates that engineers construct circuit topologies meeting functional requirements based on specific application scenarios by utilizing fundamental circuit theories and specified components. Examples include amplifier circuits, filter circuits, comparator circuits, etc.

\textbf{2) Netlist Generation}. In conventional circuit design tools, engineers manually input parameters and connectivity relationships of circuit elements to generate a textual file containing all circuit information through commercial software; this is known as a netlist. The netlist describes the type, pin connections, parameter settings, and more for each component in the circuit, serving as the foundation for subsequent design processes. However, forming a netlist directly from circuit images poses a perception-and-reasoning challenge for automated algorithms, necessitating accurate localization, classification, and topological inference. Formally, this task can be expressed as:
\begin{equation}
P_{\text{netlist}} = \mathbb{P}\left( \bigcap_{i=1}^N \{\hat{c}_i = c_i\} \cap \bigcap_{(i,j)\in\mathcal{E}} \{\hat{e}_{ij} = e_{ij}\} \right),
\end{equation}
where \(N\) denotes the total number of components in the image; \(c_i\) and \(\hat{c}_i\) represent the ground-truth and predicted class labels of the \(i\)-th component, respectively; \(\mathcal{E}\) denotes the set of true inter-component electrical connections; \(e_{ij} \in \{0,1\}\) indicates the presence \((1)\) or absence \((0)\) of a physical connection between components \(i\) and \(j\), with \(\hat{e}_{ij}\) being its predicted counterpart; and \(P_{\text{netlist}}\) quantifies the probability of perfectly recovering the complete netlist—i.e., the joint likelihood of correctly identifying all component types and accurately reconstructing all pairwise connectivity relationships simultaneously. Recent studies have also explored netlist generation. For instance, \citet{PAN2026113035} proposed a method to infer missing component connectivity from incomplete netlists, thereby partially alleviating reliance on conventional circuit design tools.

\textbf{3) Device Sizing}. In analog IC design, optimizing device dimensions is crucial for achieving the desired circuit performance. For a given topology, determining optimal parameters for active and passive components (such as the width-to-length ratio (W/L) of MOSFETs, resistances, capacitances, etc.) is required to meet specifications like gain, bandwidth, power consumption, area, common-mode rejection ratio (CMRR), among others. These objectives are often interconnected or conflicting, and as the number of devices increases, the search space for parameters expands exponentially, making manual tuning both labor-intensive and less likely to achieve optimality. 

Let \(\mathbf{x} \in \mathcal{X} \subseteq \mathbb{R}^d\) denote the vector of design parameters for circuit C, and let \(\mathbf{p}=(p_1,\dots,p_m) \in \mathcal{P}\) represent the performance metrics obtained through simulation, where \(p_i(\mathbf{x})\) is a function of \(\mathbf{x}\). The design goals are typically expressed through a figure-of-merit (FOM), which aggregates individual metrics into a composite function:
\begin{equation}
\text{FOM}(\mathbf{x}) = f\big(p_1(\mathbf{x}), \dots, p_m(\mathbf{x})\big)
\end{equation}

The optimization task aims to find the global maximum point
\begin{equation}
\mathbf{x}^* = \arg\max_{\mathbf{x} \in \mathcal{X}} \text{FOM}(\mathbf{x})
\end{equation}
or a feasible solution set satisfying all constraints. Given the strong nonlinearity, complex trade-offs, and coupling effects in analog circuits, this sizing optimization problem is inherently high-dimensional, non-convex, and computationally expensive.

To address this issue, Bayesian Optimization (BO) is widely employed in automated analog IC sizing due to its sample efficiency and ability to balance exploration and exploitation. In our study, we first utilize a parameter search agent\ref{Parameter Search Agent} to significantly reduce the original parameter space by eliminating infeasible or redundant regions based on domain knowledge and preliminary analysis. Then, we apply BO on the reduced space to identify high-quality device parameters, ensuring stable circuit operation while guaranteeing convergence reliability and minimizing costly simulations.

\textbf{4) Placement}. In analog IC design, device placement is vital for performance, matching accuracy, and robustness against process/environmental variations. Unlike digital circuits, analog designs are highly sensitive to layout effects such as device positioning, parasitics, thermal gradients, and wiring symmetry. Therefore, placement strategies must comply with geometric design rules and maintain functional integrity under manufacturing and operational fluctuations. After completing device sizing via BO, we adopt the SA algorithm for analog placement. SA balances global exploration and local refinement by adjusting the acceptance of suboptimal moves through a temperature image, effectively generating high-quality, manufacturable layouts that meet performance requirements.

\textbf{5) Routing}. Analog routing is a crucial step that bridges the SPICE netlist to physical implementation, significantly influencing analog IC performance. Unlike digital routing, which prioritizes connectivity and timing, analog routing must address non-ideal physical effects—including interconnect parasitics and device geometric mismatches—that degrade key metrics like DC gain, CMRR, bandwidth, and noise. In our framework, after performing BO-based device sizing and SA-driven placement, we conduct detailed routing using an enhanced A* search algorithm. This algorithm employs a customized heuristic cost function integrating obstacle avoidance, electrical sensitivity, and symmetry constraints, embedding analog-specific design rules into pathfinding to generate high-quality, manufacturing-compliant routes while preserving circuit electrical integrity.

\label{sec:preliminaries}

\section{method}
\subsection{Method OverView}
\begin{figure}[htbp]
\centering
\includegraphics[width=0.8\textwidth]{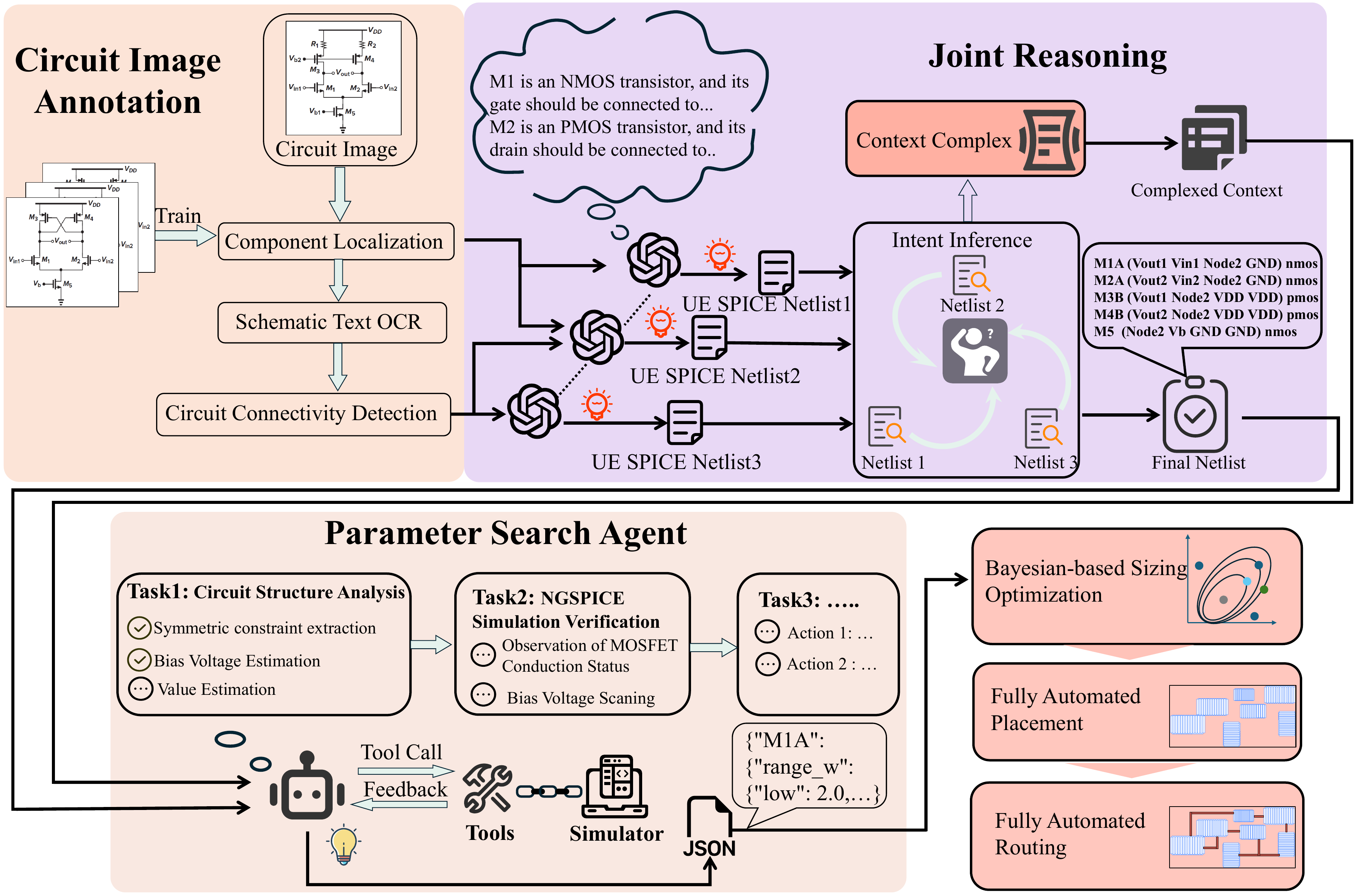}
\caption{\textbf{Workflows of AnalogMaster}: The circuit image, preprocessed via YOLO-based detection and connectivity analysis, is converted into a netlist using the proposed joint reasoning mechanism. A parameter search agent compresses the device parameter space, followed by BO for sizing, SA for placement, and A* for routing—enabling a fully automated design flow.}
\label{fig:overview}
\end{figure}

Unlike the end-to-end systems adopted in most digital IC design, analog IC design features a low level of abstraction, a limited number of standardized modular units, and strong coupling correlations between various stages in the design flow. Therefore, it is difficult to construct an end-to-end design system simply through a sequential scripting process. We present AnalogMaster, an end-to-end, training-free framework that automates the complete analog IC design flow, from circuit image input to physical layout generation. As illustrated in Figure\ref{fig:overview}, AnalogMaster integrates multiple tightly coupled stages, including circuit image preprocessing, parameterized netlist extraction, device sizing, placement, and routing. Specifically, circuit image preprocessing combines a customized YOLO-based detector, EasyOCR, and electrical connectivity analysis to recover circuit components and their interconnections. The extraction of the netlist from the image is then performed through a joint reasoning mechanism. Device sizing is formulated as a constrained optimization problem and solved using BO, while placement and routing are carried out using SA and A*-based path search, respectively. To improve optimization efficiency, a parameter search agent first analyzes the inferred circuit topology to compress the feasible design space, thereby enabling BO to more effectively explore device parameters under target performance specifications. Through the seamless integration of vision, reasoning, and optimization modules, AnalogMaster achieves accurate, scalable, and fully automated analog IC design without model retraining.

\subsection{Circuit Image Annotation}
\label{schematic annotation}
\begin{figure}[htbp]
\centering
\includegraphics[width=0.53\textwidth]{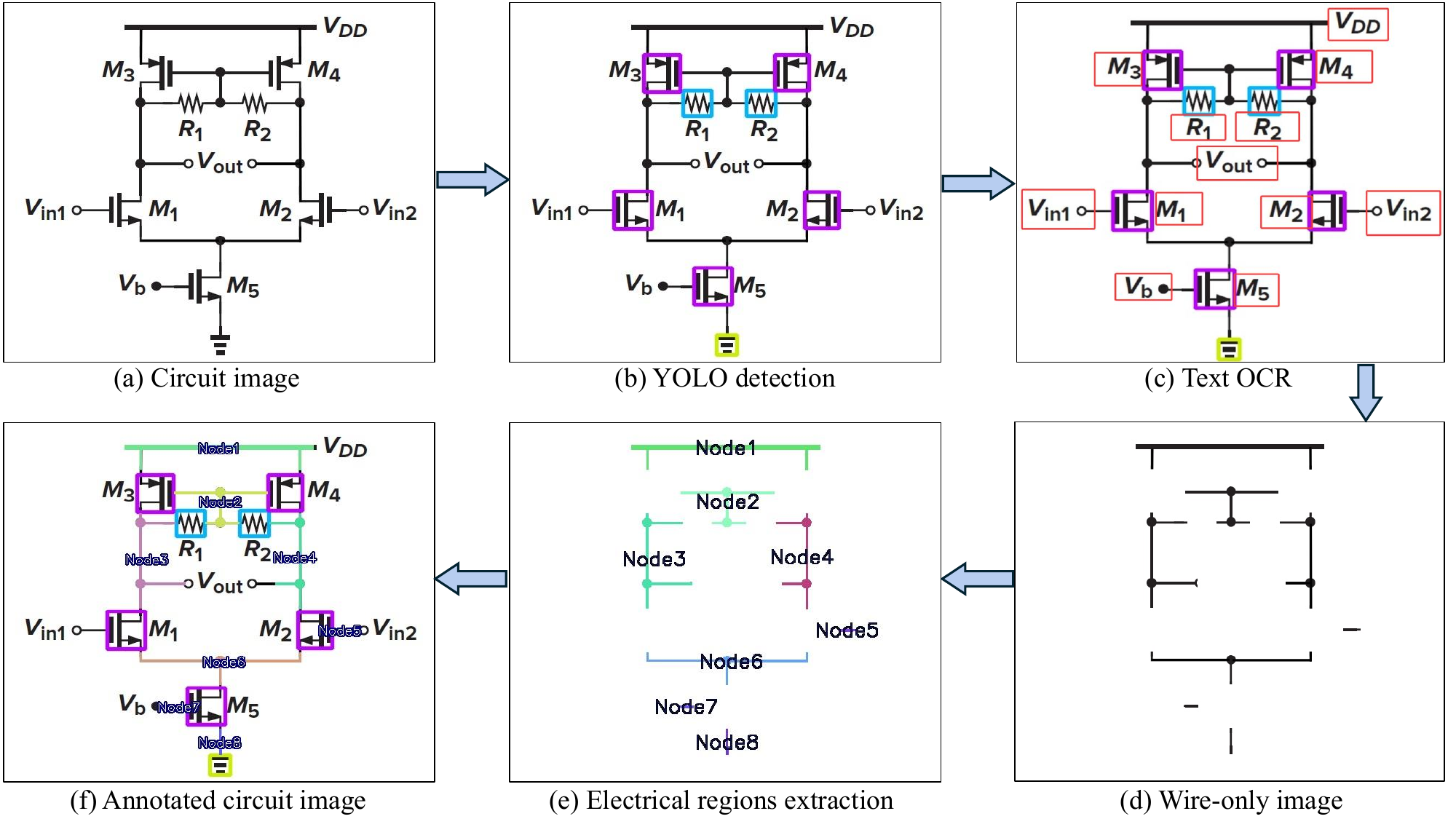}
\caption{The process flow for circuit image analysis. (a) Original circuit image. (b) YOLO-based detection of components in the circuit. (c) Text detection using OCR to identify labeled components. (d) Wire-only image extracted from the circuit. (e) Electrical regions extraction showing nodes and their connections. (f) Annotated circuit image with labeled nodes and components.}
\label{fig:img_pinline}
\end{figure}

To support robust circuit image understanding, we have meticulously constructed a dataset of approximately 10,000 high-quality annotated circuit image, covering common component categories found in analog IC schematics. Based on this dataset, we trained a YOLOv9 model specifically for circuit component detection. As shown in Figure\ref{fig:img_pinline}, the trained YOLOv9 model localizes all circuit components in the input image and outputs bounding boxes defining device positions. In parallel, EasyOCR is applied to identify and mask textual annotations, preventing interference with electrical connectivity inference. Detected component regions are also masked, yielding a wire-only image that retains only conductive paths. Morphological operations and connected-component analysis are then applied to extract electrically connected regions. Small components below a predefined area threshold, which typically correspond to non-functional ones, are discarded. Each valid connected region is assigned a unique color for visual disambiguation. To ensure consistent net representation, an overlap-aware filtering strategy based on centroid Euclidean distance is employed to merge spurious node annotations within the same electrical region. The final output integrates three complementary sources of information: YOLO-derived component bounding boxes, color-encoded connectivity regions, and refined network identifiers. This structured, multi-dimensional representation preserves both topological and semantic information, forming a robust foundation for subsequent netlist extraction.

\subsection{Joint Reasoning}
\begin{figure}[ht]
\centering
\includegraphics[width=0.8\textwidth]{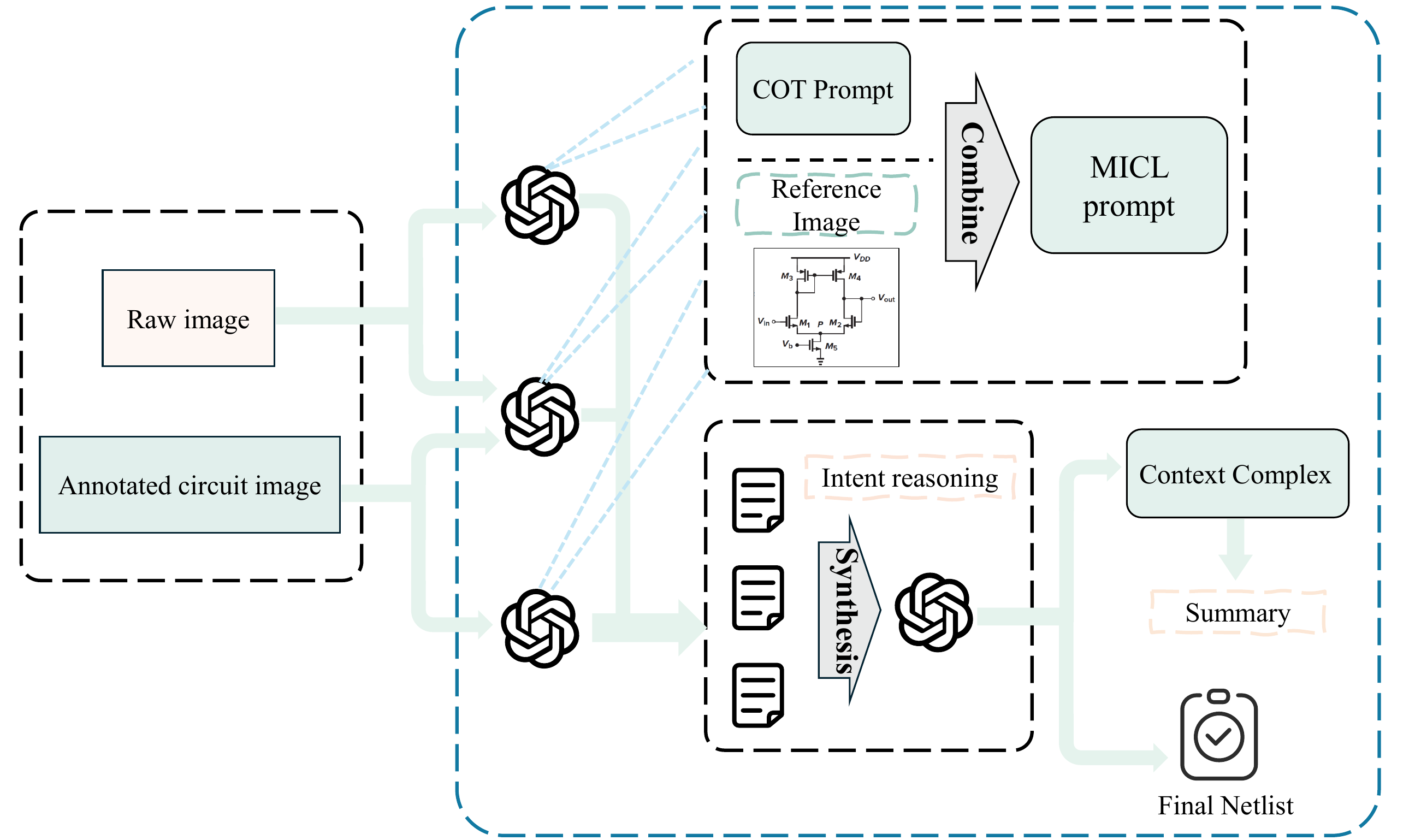}
\caption{Overview of the Joint Reasoning workflow. Three heterogeneous inputs are processed via parallel MLLM reasoning branches, followed by multi-branch result synthesis to generate accurate netlists. Compact contextual information is further extracted to facilitate downstream Bayesian optimization.}
\label{fig:JointReasoning}
\end{figure}

To improve the accuracy of circuit image-to-netlist conversion, we introduce Joint Reasoning, a training-free paradigm that enhances the image comprehension capability of general-purpose MLLMs. As illustrated in Figure \ref{fig:JointReasoning}, this framework constructs three distinct parallel inputs from raw and annotated circuit images, and feeds them into parallel MLLM branches for independent reasoning. A high-level LLM is subsequently adopted to synthesize multi-branch outputs via intent reasoning, thereby generating a reliable and accurate final netlist. Moreover, the reasoning process and circuit structural features are compressed into compact context information, which effectively narrows the parameter search space for Bayesian optimization in the downstream Parameter Search Agent.

\subsubsection{Chain-of-Thought Prompting}
\label{cot_method}
\begin{figure}[htbp]
\centering
\includegraphics[width=0.9\textwidth]{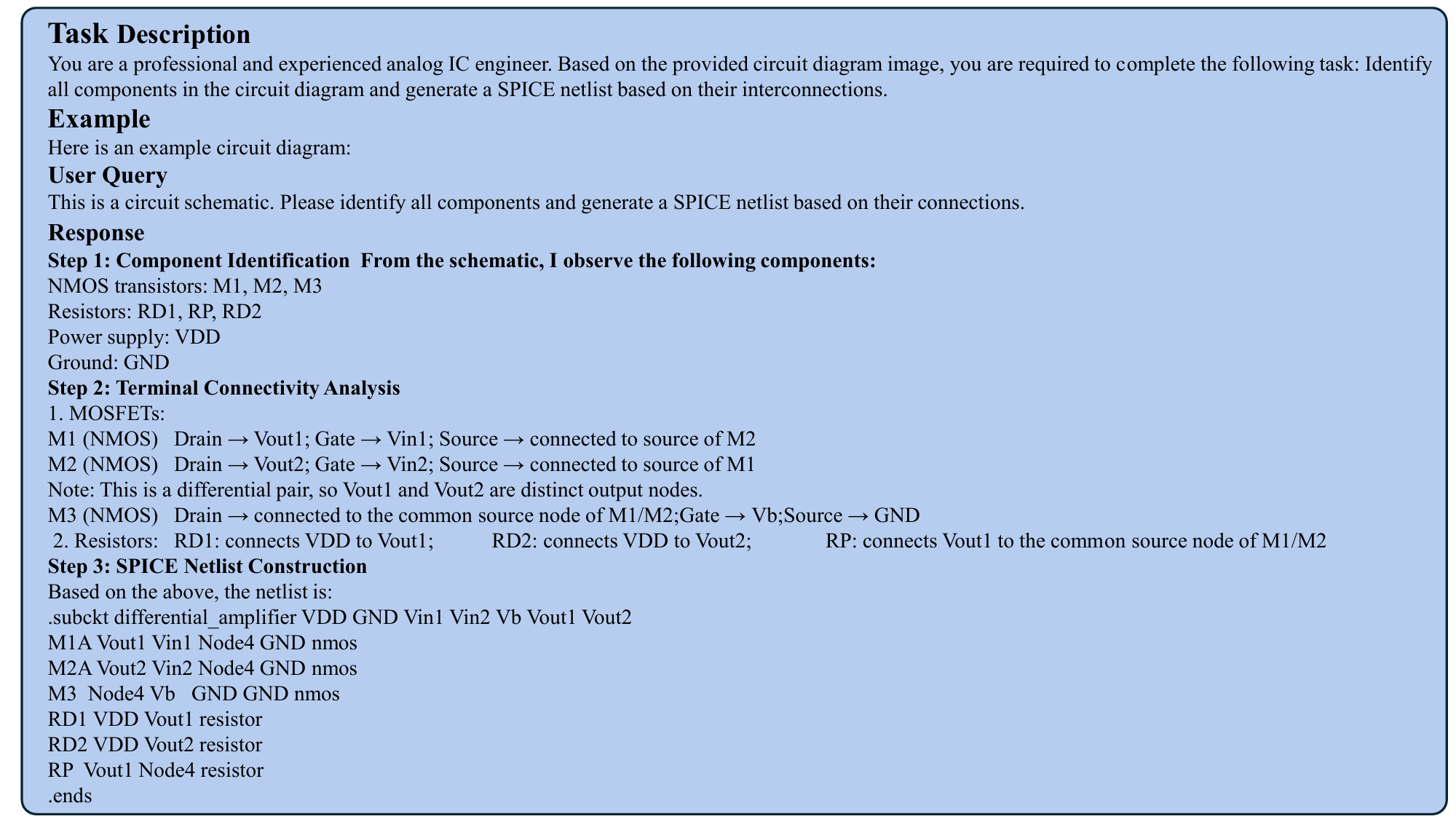}
\caption{A prompt guiding the MLLM to follow the CoT approach for sequentially analyzing a circuit image and extracting a netlist. The diagram illustrates the step-by-step process the model follows to identify components, analyze terminal connectivity, and generate the netlist from the circuit image.
}
\label{fig:cot}
\end{figure}

The conversion of a circuit images into a SPICE netlist requires three essential reasoning steps: (1) identifying all circuit components and correctly classifying their types;   (2)  determining the electrical connectivity of each terminal (port) associated with every component; and (3) merging equipotential nodes to form a structurally consistent and functionally complete netlist. To guide the MLLM through this structured reasoning process, we adopt a Chain-of-Thought (CoT) prompting strategy. As illustrated in Figure \ref{fig:cot}, the prompt first directs the model to analyze each component independently, focusing on its semantic category and the local connectivity of its ports. After completing these per-component analyses, the model is then directed to integrate the localized observations into a global circuit representation by resolving node equivalences and consolidating connections into a unified netlist. 

This stepwise decomposition not only mirrors the intrinsic logic of netlist generation but also provides flexibility in ambiguous or uncertain inference scenarios. By decoupling perception from topological reasoning, the CoT-based Joint Reasoning mechanism improves both accuracy and robustness of the final output, enabling reliable recovery of circuit semantics even in the presence of visual noise, partial occlusion, or incomplete schematic information.

\subsubsection {Multimodal In-Context Learning with Circuit Image-Driven System Prompt}

\begin{figure}[htbp]
\centering
\includegraphics[width=0.7\textwidth]{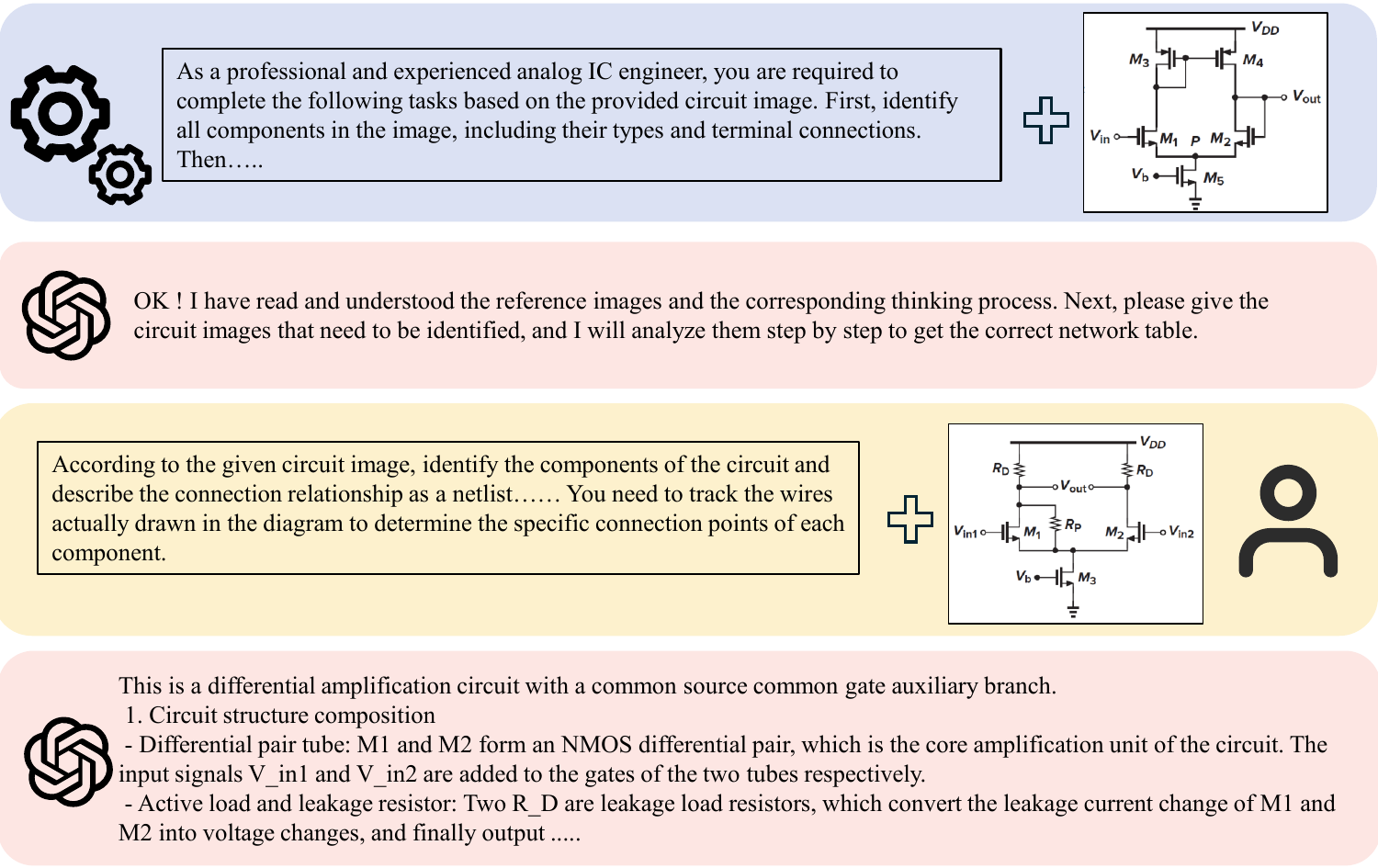}
\caption{Illustration of the multimodal context learning process. First, reference images and CoT prompts are input into the MLLM to establish a recognition logic based on the circuit image and the corresponding analysis process. Subsequently, the circuit image to be analyzed is provided to the model for netlist extraction.}
\label{fig:ICL}
\end{figure}

While the CoT prompting effectively enhances the MLLM’s logical reasoning capability in recognizing circuit diagrams and converting them into netlists, it fails to directly provide specific examples for the model to learn from. Different from pure text-based system prompts that lack multimodal guidance, we introduce the circuit image–driven multimodal in-context learning (MICL) strategy to make up for this deficiency, further improving the accuracy and generalization of netlist generation. As illustrated in Figure \ref{fig:ICL}, we first select a representative reference circuit image and pair it with a carefully designed CoT prompt that explicitly captures the complete reasoning trajectory, spanning component recognition, connectivity inference, and netlist synthesis.  This multimodal input is then provided to the MLLM, and the entire model response, including intermediate reasoning steps and the final model output, is preserved verbatim. The recorded interaction serves as a contextual anchor and is prepended to subsequent inference queries. By conditioning new inputs on this exemplar-driven context, the model is guided to reproduce a consistent, structured, and semantically grounded reasoning process across diverse circuit images. In effect, MICL emulates the role of a multimodal system prompt, enabling reliable few-shot adaptation without requiring architectural modification, additional supervision, or model fine-tuning.

\subsubsection{Parallel Reasoning}

Due to the inherent stochasticity of MLLM decoding, repeated inferences using identical CoT prompts and MICL contexts may yield slight output variations. Such inconsistencies can result in partial netlist extraction errors, which may propagate to downstream design stages and compromise the overall automation flow.  To enhance robustness and mitigate this instability, we introduce a parallel reasoning mechanism that exploits multiple complementary visual representations of the input circuit. As shown in figure\ref{fig:overview}, building upon the circuit image annotation pipeline described in \ref{schematic annotation}, the reasoning process is decomposed into three parallel branches: (1) the original raw circuit image; (2) a structurally enhanced circuit image derived from electrical connectivity analysis, in which components and wiring are disentangled and annotated; and (3) a dual-images input that jointly presents both representations, enabling cross-view comparison and collaborative reasoning. These branches operate in a decoupled and parallel manner during inference, without introducing additional computational passes beyond a single inference step. This multi-perspective reasoning strategy substantially improves accuracy, consistency, and fault tolerance, effectively suppressing stochastic decoding artifacts and enhancing the reliability of netlist extraction.

\subsubsection{Intent reasoning}
\label{intent reasoning}
Given the high complexity of converting analog IC circuit diagrams to SPICE netlists, even with the integration of CoT and MICL mechanisms, inherent limitations persist. As LLMs are essentially probabilistic models and the task itself involves intricate circuit topology analysis and electrical connectivity verification, partial errors may still occur during the diagram-to-netlist conversion process. Notably, minor errors in this process can lead to catastrophic failures in the entire circuit system, which is unacceptable in an end-to-end integrated system. 
\begin{figure}[ht]
\centering
\includegraphics[width=0.6\textwidth]{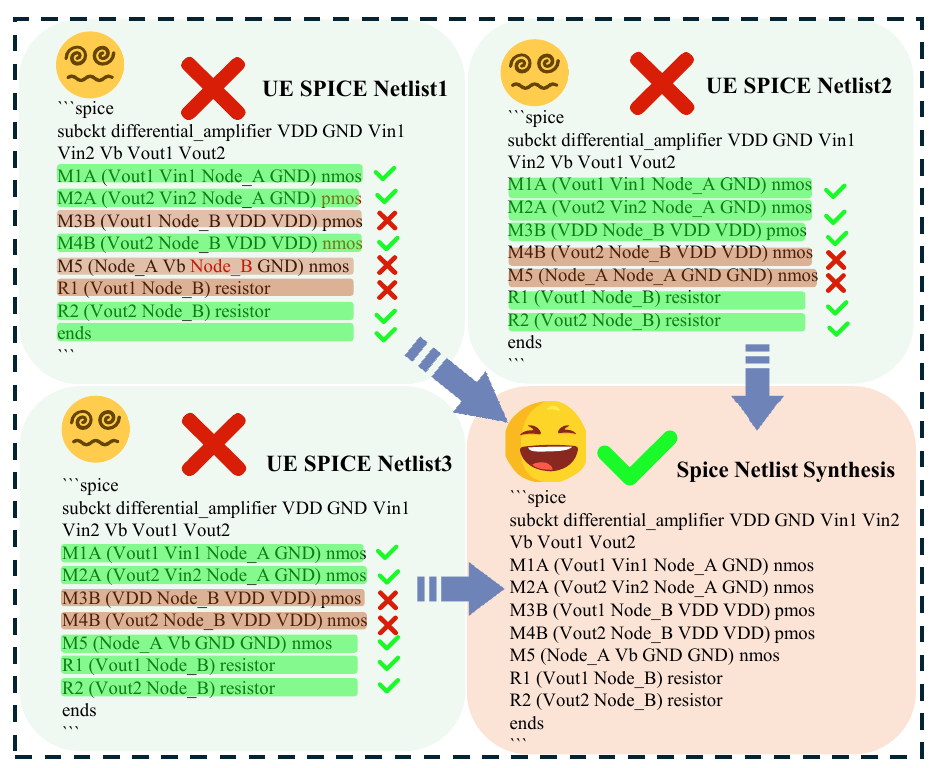}
\caption{Intent reasoning process for netlist synthesis, which demonstrates how intent reasoning extracts the correct components from three incorrect netlist results and combines them to form a fully accurate netlist.}
\label{fig:Intent}
\end{figure}

Therefore, as illustrated in Figure \ref{fig:Intent}, we introduce a text-generation LLM as a high-level inference engine that synchronously consumes the outputs of the three aforementioned parallel multimodal reasoning branches from the Parallel Reasoning mechanism. It should be emphasized that these three reasoning branches are completely independent, even though they adopt the same model in the Parallel Reasoning. This independence stems from the fact that the model receives entirely different inputs in each branch, both in terms of input quantity and content. Specifically, as elaborated in the Parallel Reasoning section, the three branches correspond to three distinct visual representations of the input circuit. The output from each reasoning branch is treated as a prior hypothesis of the circuit structure. Owing to their distinct perspectives, including component localization, electrical connectivity, and cross-view comparison, the error locations and failure modes produced by these branches when processing complex circuits are inherently diverse and partially uncorrelated.

Through a context-aware intent reasoning mechanism—essentially an error correction and fallback strategy—the text-generation LLM performs semantic fusion and consistency correction across these heterogeneous, complementary, and partially redundant multi-dimensional information. The LLM takes the output of each of the three independent reasoning branches as a key clue, comprehensively analyzing and cross-verifying the components and logical connections contained in each branch. On this basis, it identifies and eliminates erroneous components and incorrect logical connections that may exist in individual branches. By integrating the valid information from all branches and combining the inherent functional logic of analog circuits, the LLM infers and restores the most reasonable circuit configuration that meets the expected functional requirements. Finally, with this restored functional circuit as the target benchmark, the LLM further extracts the correct components and reliable logical connection relationships from each reasoning branch output, systematically integrating these correct parts to reconstruct the intended circuit topology. This multi-step reasoning and integration process ultimately yields a structurally consistent and semantically correct SPICE netlist that fully conforms to the design intent of the original circuit diagram.

\subsubsection{Context Compression}

\begin{figure}[h]
\centering
\includegraphics[width=0.6\textwidth]{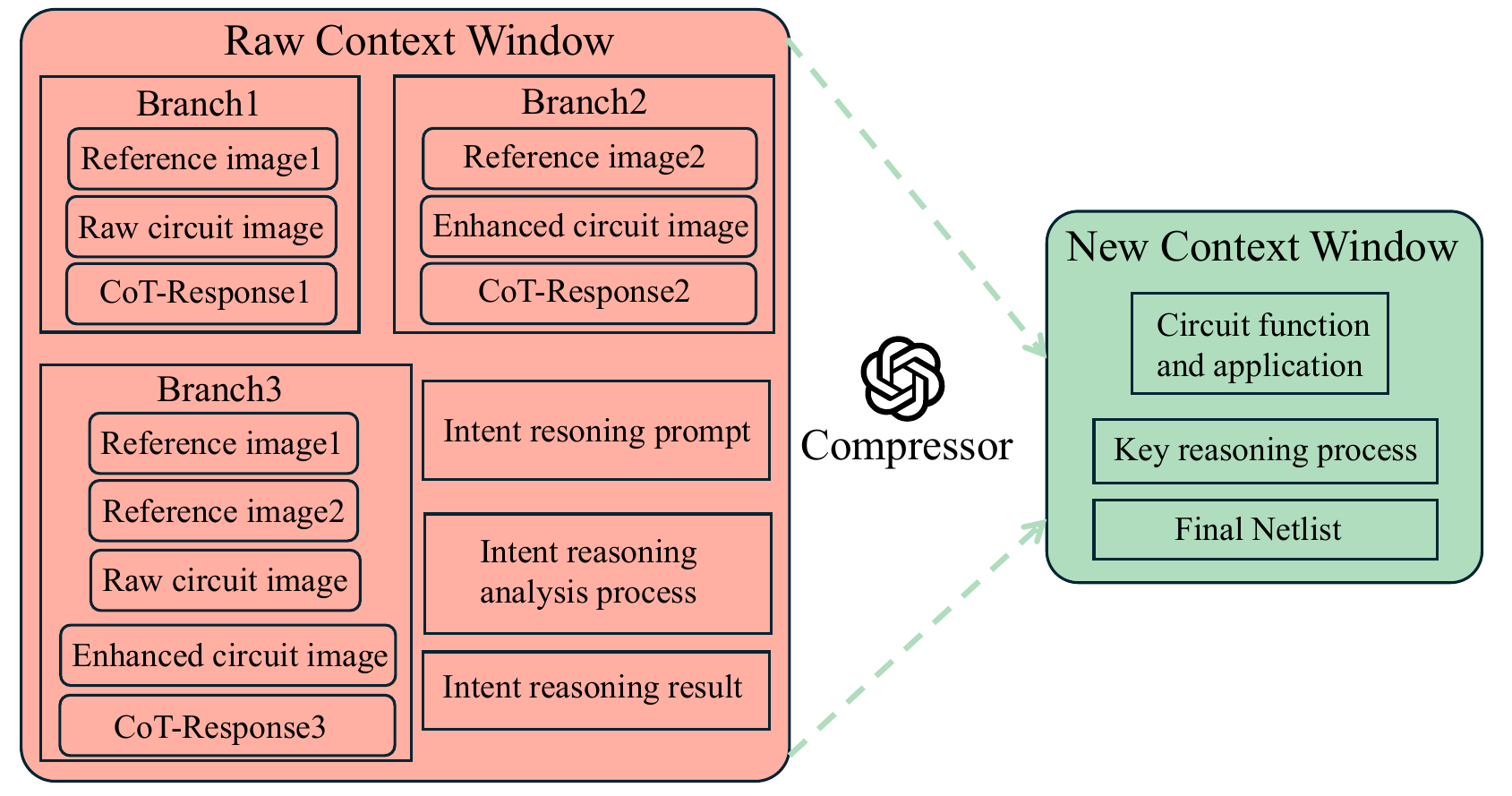}
\caption{Illustration of context compression using an LLM. The diagram shows how raw context is summarized into a compressed context, enabling more efficient processing and reasoning.}
\label{fig:context}
\end{figure}

Mainstream LLMs rely on attention mechanisms for contextual modeling\citep{vaswani2017attention},  but their quadratic computational complexity leads to performance degradation when processing long sequences. In the proposed framework, reasoning traces generated by MLLMs and LLMs contain valuable prior knowledge for subsequent Parameter Search Agents. However, their length and redundancy can significantly reduce decision-making efficiency if incorporated directly as contextual input. To address this challenge, as illustrated in Figure \ref{fig:context}, we propose a lightweight context compression strategy. After the text-generation LLM completes intent reasoning and produces the correct SPICE netlist, semantic extraction is simultaneously applied to the reasoning records generated by the three multimodal branches.  

This process preserves the core analytical logic and conclusions that are directly relevant to the verified circuit structure, while discarding irrelevant details and redundant intermediate steps.  The resulting compressed context substantially reduces sequence length and alleviates the computational burden of the attention mechanism. At the same time, it retains sufficient structural and semantic information to guide the Parameter Search Agent effectively, enabling more efficient and targeted exploration of the device parameter space during subsequent optimization stages.

\subsubsection{Advantages of Joint Reasoning}
\label{pass@k_joint_reasoning}
The proposed joint reasoning framework improves both the robustness and correctness of netlist extraction through two main approaches: 1) multi-view hypothesis expansion and 2) in-context distributional shaping.

Let $\mathcal{Y}$ denote the space of syntactically and electrically valid netlists, and let $y^* \in \mathcal{Y}$ be the ground-truth netlist corresponding to a given circuit image. In standard single-branch inference, an MLLM with parameters $\theta$ takes a fixed visual input $x$ and defines a conditional distribution $p_\theta(y \mid x)$ over candidate outputs $y \in \mathcal{Y}$. Due to stochastic decoding, each forward pass yields a random sample $y \sim p_\theta(y \mid x)$. This single sample being correct, or equal to $y^*$, has a probability that is precisely the $\text{Pass@1}$ metric:
\begin{equation}
\text{Pass@1}_{\text{single}} = \mathbb{P}_{y \sim p_\theta(\cdot \mid x)}(y = y^*) 
\end{equation}

Repeating this process $k$ times yields the conventional $\text{Pass@}k_{\text{single}} = 1 - \big(1 - \text{Pass@1}_{\text{single}}\big)^k$. However, the performance gain of repeated sampling saturates when errors across samples are correlated, which is common under identical input conditions. In contrast, the proposed joint reasoning framework employs three distinct visual encodings of the same circuit: 1) $x^{(1)}$: the original raw image,  2) $x^{(2)}$`: a structurally enhanced circuit image derived from electrical connectivity analysis, 3)$x^{(3)}$: a dual-view composition of both. Each induces a separate decoding distribution $p_\theta(y \mid x^{(b)})$ for branch $b \in \{1,2,3\}$. Notably, the three branches are statistically independent in terms of error modes—though they adopt the same MLLM, they receive entirely different inputs , as elaborated in the Intent Reasoning section, which ensures their errors are uncorrelated and lays the foundation for subsequent probability analysis. To formalize the benefit of this diversity, we define the high-probability support of branch $b$ as
\begin{equation}
\mathcal{C}^{(b)} = \big\{ y \in \mathcal{Y} \,\big|\, p_\theta(y \mid x^{(b)}) > \epsilon \big\}  
\end{equation}
where $\epsilon > 0$ is a small constant that excludes negligible-probability outliers. Intuitively, $\mathcal{C}^{(b)}$ contains all netlists that branch $b$ can plausibly generate in a typical inference. Crucially, due to input heterogeneity, we observe $\mathcal{C}^{(i)} \not\subseteq \mathcal{C}^{(j)}$ for $i \neq j$—meaning each branch covers parts of $\mathcal{Y}$ missed by others. The intent reasoning module then acts as a consensus selector over the union $\bigcup_{b=1}^3 \mathcal{C}^{(b)}$. Consequently, the success probability of joint reasoning satisfies
\begin{equation}
\text{Pass@1}_{\text{joint}} 
= \mathbb{P}\Big( y^* \in \bigcup_{b=1}^3 \mathcal{C}^{(b)} \Big)
\geq 1 - \prod_{b=1}^3 \Big(1 - \mathbb{P}(y^* \in \mathcal{C}^{(b)})\Big)
\geq \text{Pass@3}_{\text{single}}
\end{equation}
The final inequality holds because errors across branches are rarely correlated. Empirically, when one branch misidentifies a component or connection, other branches often recover the correct interpretation. This diversity ensures that joint reasoning performs at least as well as, and usually better than, three independent runs of a single-branch system.

Furthermore, the CoT prompts and MICL examples embedded in each branch serve as distributional biases. Formally, token-level generation follows
\begin{equation}
p_\theta(y_t \mid y_{<t}, x^{(b)}, c^{(b)}) \propto \exp\big( f_\theta(y_{<t}, x^{(b)}, c^{(b)}) \big)   
\end{equation}
where $c^{(b)}$ denotes the CoT/MICL context specific to branch $b$, and $f_\theta(\cdot)$ is the model’s logit function. Each $(x^{(b)}, c^{(b)})$ pair steers the autoregressive process toward reasoning paths consistent with its visual semantics.

Finally, the text-generation LLM used for intent reasoning approximates a structured consensus posterior:
\begin{equation}
p_\phi(y \mid y^{(1)}, y^{(2)}, y^{(3)}) 
\propto p_{\text{struct}}(y) \cdot \prod_{b=1}^3 \mathbb{I}\big[ \text{Consistent}(y, y^{(b)}) \big]    
\end{equation}
where $p_{\text{struct}}(y)$ encodes domain-specific constraints, and $\mathbb{I}[\cdot]$ is an indicator of semantic compatibility implicitly learned during pretraining. This fusion suppresses stochastic artifacts inconsistent across views, effectively implementing constrained decoding without explicit search.

Although the framework could be extended by introducing additional semantically equivalent visual variants to further enlarge the hypothesis space, practical limitations arise as the number of branches increases.  Additional branches expand the input context length for the intent-reasoning LLM, potentially degrading reasoning quality due to finite context windows. Moreover, marginal gains diminish as branch representations become redundant, while computational cost grows linearly. In practice, we find that three carefully designed branches (raw, structurally enhanced, and dual-view) strike a favorable balance between diversity, efficiency, and accuracy.

\subsection{Parameter Search Agent}
\label{Parameter Search Agent}

Circuit images typically lack detailed component parameter information. However, key parameters are critical for ensuring correct circuit operation and meeting target performance specifications, such as resistor values, MOSFET channel dimensions (W/L), and bias voltages. To address this challenge, we introduce an LLM-based parameter search agent. Through Self-Enhanced Prompt Engineering and a multi-tool collaborative invocation mechanism, the agent constructs a feasible and structured parameter search space, enabling stable circuit functionality and effective performance optimization.

\subsubsection{Self-Enhanced Prompt Engineering}

We propose a self-enhancing prompt engineering strategy to address the limitations of conventional ReAct agents. Standard ReAct agents suffer from local reasoning bias due to the lack of macro-level planning, causing inefficient tool usage and redundant actions. They operate in a stepwise, feedback-driven trial-and-error mode analogous to Bayesian optimization , which improves accuracy via accumulated prior knowledge but is ill-suited for LLMs. LLM inference incurs high computational costs, and iterative feedback loops expand context windows, drastically reducing efficiency. Moreover, given the excessively large parameter space in device sizing, direct initialization and execution of BO often fail to converge to feasible solutions, necessitating prior parameter space compression. Unlike BO that pursues precise solutions, LLMs excel at efficiently narrowing the parameter search space in a coarse-grained manner. In contrast, our strategy enables the agent to first conduct high-level task planning based on circuit topology, device characteristics, and design constraints, decomposing global optimization objectives into ordered, prioritized sub-tasks. Instead of completely  relying on real-time feedback, the agent follows a refined "Thought–Action–Observation" cycle to progressively narrow and refine the parameter space.

As illustrated in Figure \ref{fig:overview}, this design enables the agent to perform macro-level reasoning while sequentially completing the planned task list. By mitigating the weak long-horizon planning capability inherent in standard ReAct agents, the proposed method supports adaptive and stable parameter space construction across a wide range of circuit complexities, from simple amplifiers to multi-module analog systems.

\subsubsection{Context Truncation}
\begin{figure}[htbp]
\centering
\includegraphics[width=0.7\textwidth]{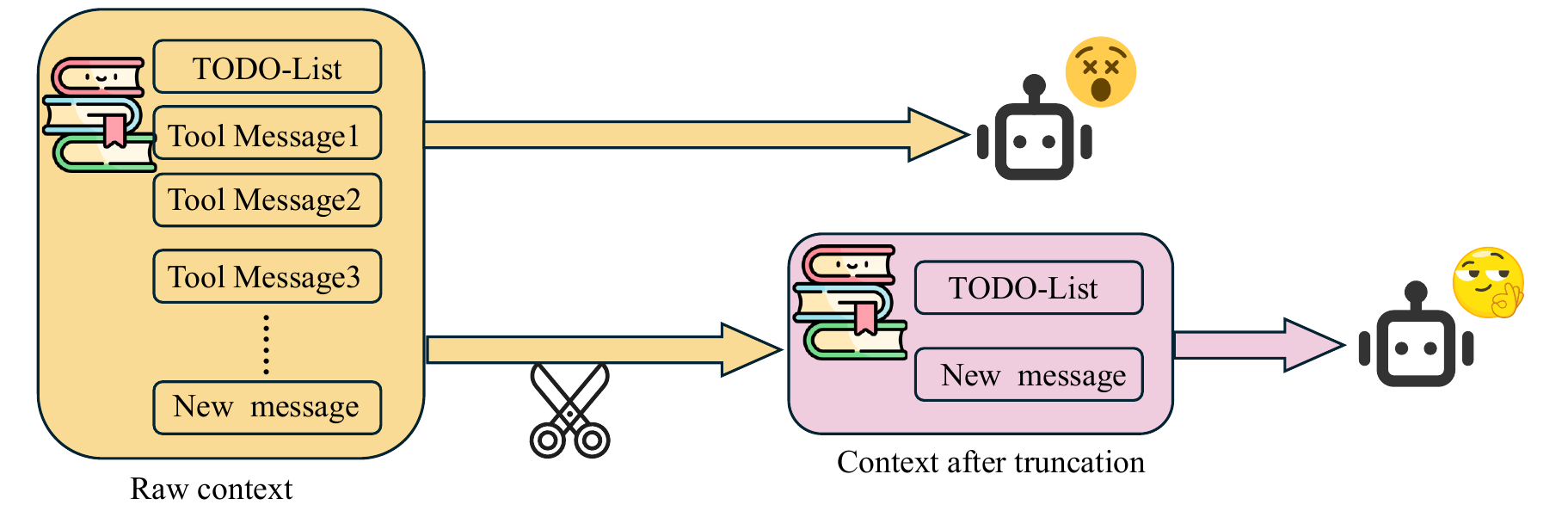}
\caption{Comparison before and after context truncation. This diagram demonstrates the process of removing redundant historical tool messages and retaining only the most relevant information, such as the TODO list and the latest tool message.}
\label{fig:truncation}
\end{figure}
During iterative tool invocation, the agent continuously accumulates feedback information, resulting in increasingly long context sequences. Excessive context length can obscure critical information and degrade the model’s ability to retain the high-level planning objectives specified in the to-do list. Rather than applying generic context compression techniques, we introduce a context truncation mechanism tailored specifically for agent-based parameter search. A critical observation is that the specific information returned by tools (e.g., MOSFET operating state parameters) is valid for the current parameter configuration. When the agent transitions to a new parameter candidate, these tool outputs must be regenerated and therefore, do not need to be retained. Exploiting this property, the context truncation mechanism selectively discards obsolete tool feedback while preserving the agent’s planning state and decision logic. As shown in figure\ref{fig:truncation}, this significantly reduces token computational overhead by eliminating redundant information, while ensuring that the agent maintains access to valid context. By balancing efficiency and accuracy, context truncation provides robust support for continuous, iterative parameter search.

\label{sec:method}

\section{Experiment}

\subsection{Dataset}

We construct the Circuit Element Detection (CED) dataset, a comprehensive collection meticulously aggregated from multiple public resources. Specifically, the original dataset is composed of 3,334 images from \citep{CNN_CircuitComputerVisionModel}, 1,613 images from \citep{MOSFETDetectionComputerVisionDataset}, and 734 images from \citep{prokComputerVisionDataset}, summing up to 5,681 raw images. We eliminated samples representing irrelevant contexts that would not be encountered in practical applications, thereby ensuring the dataset accurately reflects realistic, high-quality circuit image environments. To enrich the data diversity and improve model generalization, we apply a set of effective data augmentation strategies, including rotation, random brightness adjustment, and random cropping, which effectively expands the dataset to the final scale of 9,753 images. The dataset is systematically partitioned into training, validation, and test sets with an 8:1:1 ratio to ensure a balanced evaluation framework.

\begin{figure}[htbp]
\centering
\includegraphics[width=0.9\textwidth]{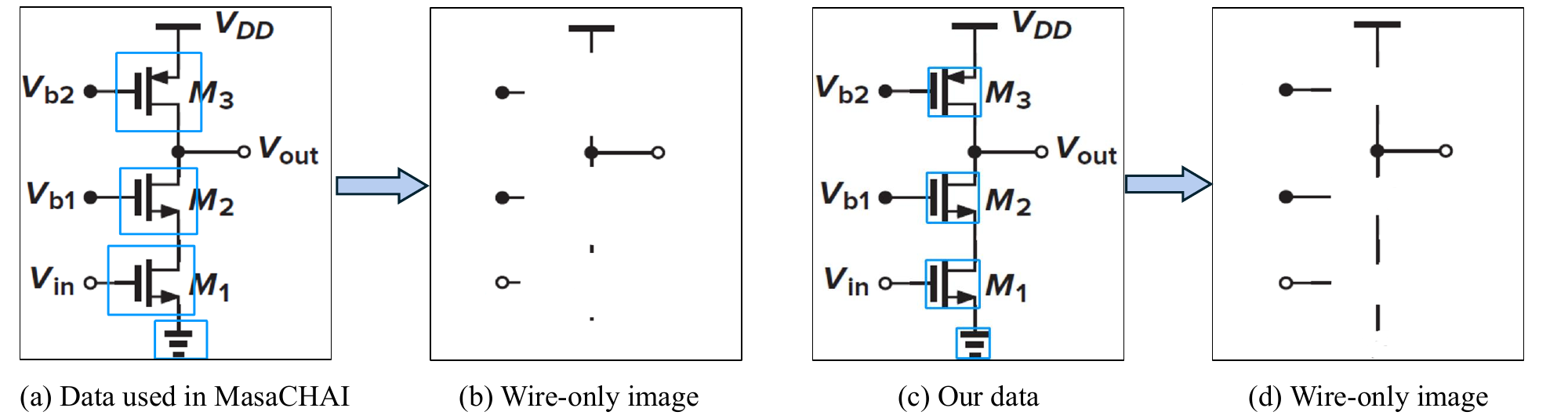}
\caption{Comparison of circuit element detection in different datasets. (a) Circuit image from the dataset used in MasaCHAI\citep{masalaCHAI-2025}. (b) Wire-only image derived from the dataset in (a). (c) Circuit image from our dataset. (d) Wire-only image derived from our dataset.}
\label{fig:yolo_compare}
\end{figure}

The CED dataset encompasses 12 fundamental circuit element classes, including AC Source, BJT, Battery, Capacitor, DC Source, Diode, Ground, Inductor, MOSFET, Resistor, Current Source, and Voltage Source, which collectively constitute the building blocks of analog integrated circuits. As depicted in Figure \ref{fig:yolo_compare}, compared with the v3qwe\citep{v3qwe} dataset adopted by MasaCHAI, our dataset facilitates more accurate detection of circuit elements, significantly reducing bounding box offset errors. This improvement directly enhances wire continuity preservation in subsequent connectivity inference.

\subsection{Experimental Environment}

All experiments were conducted on a unified computing platform to ensure consistency and reproducibility. The hardware configuration consists of an Intel(R) Xeon(R) Silver 4310 processor with a base frequency of 2.10 GHz, 282 GB of memory, and an NVIDIA A100-SXM4-40GB Graphics Processing Unit (GPU). The process library for all circuits is open-sourced SKY130PDK. The proposed framework was implemented using the LangGraph framework. Supporting tools and libraries include PyTorch (2.5.1), ngspice (43) for circuit simulation, and OpenCV-Python (4.11.0.86) for image processing. The training hyperparameters  used for YOLOv9 are summarized in Table \ref{tab:YOLO}.

\begin{table}[htbp]
\centering
\caption{Parameter of traing YOLO}
\label{tab:YOLO}
\begin{tabular}{ll}
\hline

\rule{0pt}{10pt}\textbf{Parameter} & \textbf{Value} \rule[-5pt]{0pt}{0pt} \\ \hline  
Epochs                   & 400       \\
Training Batch Size      & 32        \\
Testing Batch Size       & 32        \\
Image Size               & 640 × 640 \\
Number of Worker Threads & 8         \\
Optimizer                & SGD     \\
IflrAuto                 & True      \\
Initial Learning Rate    & 0.001     \\
Patience                 & 100       \\
Random Seed for Reproducibility   &0 \\
Weight Decay            &0.0005      \\
Momentum                 & 0.937     \\  \hline
\end{tabular}
\end{table}

\subsection{Benchmark}

Following the evaluation paradigm established by AnalogCoder\citep{lai2025analogcoder}, we selected 15 representative circuit images from the AnalogGenies\citep{gao2025analoggenie} dataset as test benchmarks, as shown in Table \ref{tab:task_description}. These circuits were classified into three difficulty levels: simple, medium, and difficult. This classification is based on factors such as the number of components, the degree of image interference, and the complexity of the interconnection topology.
\begin{table}[htbp]
  \centering
  \caption{\textbf{Benchmark Description}. All circuit task descriptions with varying difficulty levels are listed herein. The difficulty is categorized into three levels:\textbf{\textcolor[HTML]{afddbe}{Easy}}, \textbf{\textcolor[HTML]{b7cdf0}{Medium}}, and \textbf{\textcolor[HTML]{f1cbde}{Difficult}}.} 
  \label{tab:task_description} 
  \small 
  \begin{tabular}{c|l}
    \toprule 
    \textbf{Case} & \textbf{Circuit Description}                                                                                      \\ 
    \midrule 
    \rowcolor[HTML]{afddbe} 
    1           & NMOS-input cascode amplifier                                                                              \\
    \rowcolor[HTML]{afddbe} 
    2           & CMOS differential amplifier with common-mode rejection            \\
    \rowcolor[HTML]{afddbe} 
    3           & CMOS differential pair with active load                                                                               \\
    \rowcolor[HTML]{afddbe} 
    4           & NMOS cross-coupled cascode  amplifier                                   \\
    \rowcolor[HTML]{b7cdf0} 
    5           & Differential pair with source degeneration                                                            \\
    \rowcolor[HTML]{b7cdf0}
    6           & CMOS differential amplifier with source degeneration                                              \\
    \rowcolor[HTML]{b7cdf0} 
    7           & CMOS differential amplifier with current-source load                         \\
    \rowcolor[HTML]{b7cdf0} 
   8           & Differential pair with active load                                                                            \\
     
    \rowcolor[HTML]{b7cdf0} 
    9           & CMOS differential amplifier with resistive feedback                                                           \\
    \rowcolor[HTML]{b7cdf0} 
    10          & Feedback-enhanced NMOS cascode amplifier                                                                      \\
    \rowcolor[HTML]{f1cbde} 
    11          & Two-stage cascaded differential amplifier                                                                     \\
    \rowcolor[HTML]{f1cbde} 
    12          & Differential pair with active load and cascode configuration                                                 \\
    \rowcolor[HTML]{f1cbde} 
    13          & Multistage CMOS operational amplifier                                                                                     \\
    \rowcolor[HTML]{f1cbde}
    14          & Dynamic Latch Comparator
                                                \\
    \rowcolor[HTML]{f1cbde} 
    15          & Differential pair with active load and cascode structure                                                                  \\
    \bottomrule 
  \end{tabular}
\end{table}
For the evaluation process to be considered successful, the proposed framework must satisfy the following criteria simultaneously: 1. Accurately identify the input circuit image and generate the correct netlist; 2. Successfully optimize device parameters to meet the target circuit performance specifications; 3. Ensure the placement and routing results comply with process design constraints.

\subsection{Evaluation Metric}
\label{pass@k}
We adopt Pass@k\citep{chen2021evaluating}, which provides an unbiased approximation of the Pass@k defined in our analysis\ref{pass@k_joint_reasoning}, as the primary evaluation criterion. Pass@k measures the probability that at least one valid result satisfying the task requirements is produced within a maximum of k independent generation attempts. Compared with single-generation accuracy, Pass@k captures the potential performance ceiling of the model in scenarios involving randomness or uncertainty (e.g., non-deterministic decoding in LLMs). By allowing multiple samples, Pass@k provides a more comprehensive evaluation of model performance. The mathematical definition is as follows:

\begin{equation}
\text{Pass@}k = 1 - \frac{\binom{n - c}{k}}{\binom{n}{k}} 
\end{equation}

where \(n\) represents the total number of attempts, \(c\) represents the number of successful generations, and \(\binom{n}{k}\) is the binomial coefficient. In all experiments, we set \(n\) = 15. We conducted a systematic evaluation of four MLLMs: Qwen-VL-Max, GLM-4.5V, GPT-4o-mini, and GPT-5.

\subsection{Bayesian Optimization Experimental Setup}
To achieve automatic parameter tuning of the circuit under test, a Bayesian optimization framework is employed as the core experimental precondition for circuit performance optimization. The Tree-structured Parzen Estimator (TPE) is adopted as the surrogate model, and the Expected Improvement (EI) is utilized as the acquisition function to balance the exploration of untested parameter regions and the exploitation of high-performance regions in the parameter space. The parameter search space is automatically determined by the parameter search agent after a limited number of iterations. A total of 100 SPICE simulation iterations is executed to balance optimization efficiency and computational cost. The optimization initialization adopts random sampling, and a median pruning strategy is integrated to early terminate unpromising trials that perform worse than the median of historical results, with convergence defined by completing a fixed number of optimization trials. The optimization direction is dynamically specified according to the target circuit performance metric, and the entire simulation process is conducted under a fixed temperature of 25℃ using the tt (typical-nmos/typical-pmos) process corner of the Sky130 PDK.

\subsection{Experimental Results}

\begin{table}[htbp]
\centering 
\caption{\textbf{Main Results}. We evaluated the task completion performance of the AnalogMaster framework using four models, namely Qwen-VL-Max, GLM-4.5V, GPT-4o-mini, and GPT-5, respectively.} 
\label{tab:Main_Result} 
\small 
\begin{tabular}{c|cc|cc|cc|cc}
\toprule 
\multirow{2}{*}{\textbf{Case}} & \multicolumn{2}{c|}{\textbf{GPT-4o-mini}} & \multicolumn{2}{c|}{\textbf{Qwen-VL-Max}} & \multicolumn{2}{c|}{\textbf{GLM-4.5V}}   & \multicolumn{2}{c}{\textbf{GPT-5}}        \\
                    & Pass@1          & Pass@5         & Pass@1          & Pass@5         & Pass@1         & Pass@5         & Pass@1         & Pass@5         \\ 
\midrule 
1       & \textbf{100.0}  & \textbf{100.0} & \textbf{100.0}  & \textbf{100.0} & \textbf{100.0} & \textbf{100.0} & \textbf{100.0} & \textbf{100.0} \\
2       & \textbf{100.0}  & \textbf{100.0} & \textbf{100.0}  & \textbf{100.0} & \textbf{100.0} & \textbf{100.0} & \textbf{100.0} & \textbf{100.0} \\
3       & 86.7            & \textbf{100.0} & 46.7            & 98.1           & \textbf{100.0} & \textbf{100.0} & \textbf{100.0} & \textbf{100.0} \\
4       & 93.3            & \textbf{100.0} & 93.3            & \textbf{100.0} & \textbf{100.0} & \textbf{100.0} & \textbf{100.0} & \textbf{100.0} \\
5       & 99.3            & \textbf{100.0} & \textbf{100.0}  & \textbf{100.0} & \textbf{100.0} & \textbf{100.0} & \textbf{100.0} & \textbf{100.0} \\
6       & 86.7            & \textbf{100.0} & 93.3            & \textbf{100.0} & \textbf{100.0} & \textbf{100.0} & \textbf{100.0} & \textbf{100.0} \\
7       & 6.7             & 33.3           & 20.0            & 73.6           & 0.0              & 0.0              & \textbf{100.0} & \textbf{100.0} \\
8       & 40.0            & 95.8           & 13.3             & 57.1           & 33.3           & 91.6           & \textbf{73.3}           & \textbf{100.0} \\
9       & 26.7            & 84.6           & 60.0            & 99.8           & 46.7           & 98.1           & \textbf{100.0} & \textbf{100.0} \\
10      & 46.7            & 98.1           & 33.3            & 91.6           & 53.3           & 99.3           & \textbf{100.0}          & \textbf{100.0} \\
11      & 20.0            & 73.6          & \textbf{93.3}            & \textbf{100.0}          & 53.3           & 99.3           & 80.0           & \textbf{100.0} \\
12      & 40.0            & 95.8           & 33.3            & 91.6           & 33.3           & 91.6           & \textbf{60.0}           & \textbf{99.8}  \\
13      & 0.0             & 0.0            & 0.0             & 0.0            & 6.7            & 33.3           & \textbf{86.7}           & \textbf{100.0} \\
14     &0.0           &0.0          &6.7            &33.3       &0.0
&0.0       & \textbf{80.0}     &\textbf{100.0}   \\

15      & 0.0             & 0.0            & 6.7             & 33.3           & 0.0            & 0.0            & \textbf{100.0}          & \textbf{100.0} \\ 
\midrule 
Avg     & 49.7            & 72.1           & 53.3            & 78.6           & 55.1           & 74.2           & \textbf{92.0}  & \textbf{99.9}  \\
Solve   & 12              & 12             & 14              & 14             & 12             & 12             & \textbf{15}    & \textbf{15}    \\ 
\bottomrule 
\end{tabular}
\end{table}

\begin{figure}[htbp]
\centering
\includegraphics[width=0.8\textwidth]{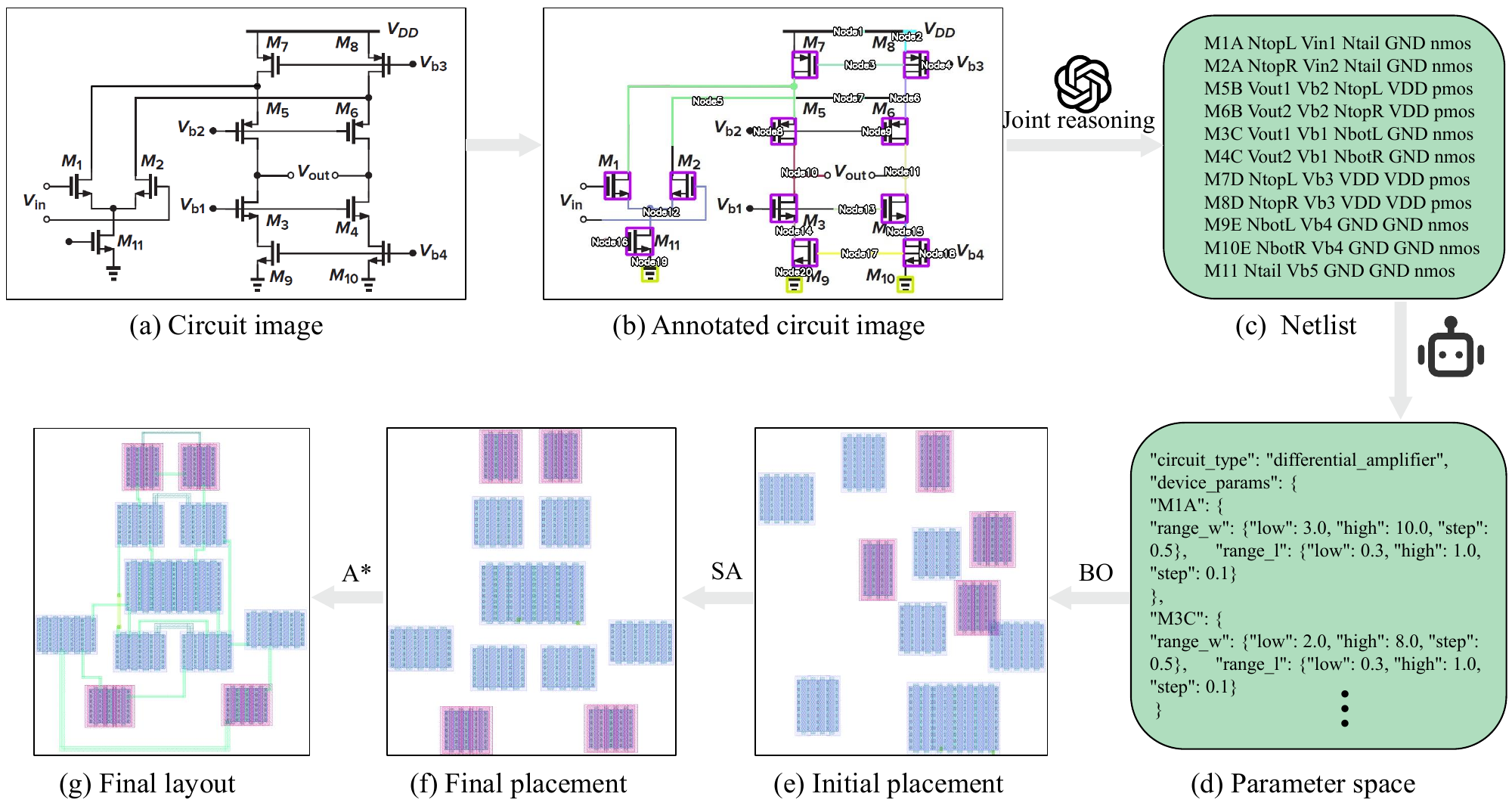}
\caption{Full execution example of the proposed AnalogMaster framework. (a) Original circuit image. (b) Annotated circuit image with identified components and nodes. (c) Generated netlist based on joint reasoning. (d) Parameter space configuration based on the agent. (e) Initial placement of components. (f) Final placement of components after SA. (g) Final layout after routing using A*.}
\label{fig:layout}
\end{figure}

Table \ref{tab:Main_Result} summarizes  the task completion performance of each MLLM within AnalogMaster under identical experimental configurations. The results indicate that the overall performance of the proposed framework is strongly influenced by the capabilities of the underlying MLLM. A complete analog IC design flow requires the model to (i) accurately identify circuit structure and generate syntactically compliant netlists, (ii) perform higher-level analytical reasoning (e.g., recognizing device symmetry and inferring circuit functionality), and (iii) adhere strictly to instruction constraints (e.g., producing SPICE netlists in a prescribed format).  Failure at any of these stages, such as generating an incorrect or noncompliant netlist, can propagate to downstream modules, including intent reasoning and the parameter search agent.  In such cases, missing or inconsistent information prevents the construction of a feasible device parameter range. The key results of the AnalogMaster framework process are shown in Figure \ref{fig:layout}. Among the four evaluated models, GPT-5 achieved the best overall performance. Moreover, the remaining three models maintain a relatively high Pass@5 success rate at k = 5 samplings, which indicates the robustness of our framework in ensuring consistent performance even when applied to different models.

However, we observed that the GPT-4o-mini, Qwen-VL-Max, and GLM-4.5V models exhibit unsatisfactory performance in certain cases. Our research reveals that this phenomenon stems from two main reasons: on the one hand, the models themselves have a certain degree of bias towards a small number of cases. Such bias causes the models to violate the objective facts of circuit images in all reasoning branches, leading to confusion between NMOS and PMOS, or arbitrary speculation about port connection relationships, which in turn results in inconsistency between the netlists and the circuit images. On the other hand, although the parameter search agent has significantly compressed the parameter search space, the BO algorithm, as a sequential optimization algorithm based on a probabilistic model, fails to meet the preset performance indicators within the effective number of iterations specified in this study. Notably, this phenomenon is significantly alleviated in GPT-5, indicating that the framework proposed in this study has certain requirements for the performance of the MLLM itself.

\subsection{Comparative Experiment}

To further evaluate the effectiveness of the proposed method, we used Qwen-VL-Max and GLM-4.6V as core models to replicate MasaCHAI\citep{masalaCHAI-2025} on the same set of 15 circuit images and calculated its Pass@k metric. While MasaCHAI performs end-to-end generation from circuit images to netlists, the proposed framework decouples netlist recognition as an independent functional module. For a fair comparison, we evaluate only the netlist recognition sub-module within our framework. Under this evaluation protocol, a test is considered successful if the model generates a SPICE netlist with a correct topological structure and compliant syntax grammar from the input image.

\begin{table}[htbp]
\centering 
\caption{Performance comparison of MasaCHAI and AnalogMaster under Qwen-VL-Max and GLM-4.6V.} 
\label{tab:comparative} 
\small 
\begin{tabular}{c|cccc|cccc}
\toprule 
\multirow{3}{*}{\textbf{Case}} & \multicolumn{4}{c|}{Qwen-VL-Max}                                                    & \multicolumn{4}{c}{GLM-4.6V}                                                       \\ \cline{2-9} 
                               & \multicolumn{2}{c|}{\textbf{MasaCHAI}} & \multicolumn{2}{c|}{\textbf{AnalogMaster}} & \multicolumn{2}{c|}{\textbf{MasaCHAI}} & \multicolumn{2}{c}{\textbf{AnalogMaster}} \\
                               & Pass@1  & \multicolumn{1}{c|}{Pass@5}  & Pass@1               & Pass@5              & Pass@1  & \multicolumn{1}{c|}{Pass@5}  & Pass@1               & Pass@5              \\ \midrule 
1                              & 6.7     & \multicolumn{1}{c|}{33.3}    & \textbf{100.0}       & \textbf{100.0}      & 0.0     & \multicolumn{1}{c|}{0.0}     & \textbf{100.0}                  & \textbf{100.0}                 \\
2                              & 33.3    & \multicolumn{1}{c|}{91.6}    & \textbf{100.0}       & \textbf{100.0}      & 86.7    & \multicolumn{1}{c|}{\textbf{100.0}}   & \textbf{100.0}                  & \textbf{100.0}                 \\
3                              & 6.7     & \multicolumn{1}{c|}{33.3}    & \textbf{53.3}        & \textbf{99.3}       & 0.0     & \multicolumn{1}{c|}{0.0}     &\textbf{100.0}                  & \textbf{100.0}                 \\
4                              & 40.0    & \multicolumn{1}{c|}{95.8}    & \textbf{100.0}       & \textbf{100.0}      & 20.0    & \multicolumn{1}{c|}{73.6}    & \textbf{100.0}                  & \textbf{100.0}                 \\
5                              & 0.0     & \multicolumn{1}{c|}{0.0}     & \textbf{100.0}       & \textbf{100.0}      & 0.0     & \multicolumn{1}{c|}{0.0}     & \textbf{100.0}                  & \textbf{100.0}                 \\
6                              & 0.0     & \multicolumn{1}{c|}{0.0}     & \textbf{100.0}       & \textbf{100.0}      & 0.0     & \multicolumn{1}{c|}{0.0}     &\textbf{73.3}                  &\textbf{100.0}                \\
7                              & 26.3    & \multicolumn{1}{c|}{84.6}    & \textbf{33.3}        & \textbf{91.6}       &\textbf{6.7}     & \multicolumn{1}{c|}{\textbf{33.3}}    & \textbf{6.7}                  & \textbf{33.3}                 \\
8                              & 6.7     & \multicolumn{1}{c|}{33.3}    & \textbf{46.7}        & \textbf{98.1}       & 6.7     & \multicolumn{1}{c|}{33.3}    & \textbf{40.0}                & \textbf{95.8}                \\
9                              & 0.0     & \multicolumn{1}{c|}{0.0}     & \textbf{66.7}        & \textbf{99.9}       & 0.0     & \multicolumn{1}{c|}{0.0}     & \textbf{46.7}                 &\textbf{98.1}                \\

10                             & 0.0     & \multicolumn{1}{c|}{0.0}     & \textbf{46.7}        & \textbf{98.1}       & 13.3    & \multicolumn{1}{c|}{57.1}    & \textbf{40.0}                  &\textbf{95.8}                \\
11                             & 0.0     & \multicolumn{1}{c|}{0.0}     & \textbf{93.3}        & \textbf{100.0}      & 0.0     & \multicolumn{1}{c|}{0.0}     & \textbf{33.3}                  & \textbf{91.6}                 \\
12                             & 0.0     & \multicolumn{1}{c|}{0.0}     & \textbf{40.0}        & \textbf{95.8}       & 0.0     & \multicolumn{1}{c|}{0.0}     & \textbf{6.7}                  &\textbf{ 33.3}                 \\
13                             & 0.0     & \multicolumn{1}{c|}{0.0}     & \textbf{6.7}         & \textbf{33.3}       & 0.0     & \multicolumn{1}{c|}{0.0}     & 0.0                  & 0.0                 \\
14                             & 0.0     & \multicolumn{1}{c|}{0.0}     & \textbf{6.7}         & \textbf{33.3}       & 0.0     & \multicolumn{1}{c|}{0.0}     & 0.0                  & 0.0                 \\
15                             & 0.0     & \multicolumn{1}{c|}{0.0}     & \textbf{6.7}         & \textbf{33.3}       & 0.0     & \multicolumn{1}{c|}{0.0}     & \textbf{6.7}                 & \textbf{33.3}                 \\ \midrule 
Avg                            & 8.0     & \multicolumn{1}{c|}{24.8}    & \textbf{60.0}        & \textbf{85.5}       & 8.9     & \multicolumn{1}{c|}{19.8}    & \textbf{50.2}                & \textbf{72.1}                 \\
Solve                          & 6       & \multicolumn{1}{c|}{6}       & \textbf{15}          & \textbf{15}         & 5       & \multicolumn{1}{c|}{5}       &\textbf{13}                 & \textbf{13}                 \\ \bottomrule 
\end{tabular}
\end{table}

As shown in Table \ref{tab:comparative}, the netlist recognition module of the proposed method significantly outperforms MasaCHAI in terms of the Pass@k metric, demonstrating superior accuracy in circuit image understanding and architectural analysis. Further analysis revealed that MasaCHAI frequently misclassifies the type of MOSFET devices and confuses source and drain terminals. Additionally, its relatively simple prompt design often produces netlists with inconsistent formatting, resulting in parsing failures or deviations from task specifications that hinder reliable information extraction. It is worth noting that the netlist recognition module evaluated here represents only a subset of the overall AnalogMaster framework. Despite this, it exhibits substantially lower parameter count and computational overhead than MasalaCHAI’s end-to-end system, highlighting its lightweight design and stronger engineering practicality.

Additionally, we conducted a focused comparison of circuit component detection performance. As shown in Table \ref{tab:component_accuracy_single}, the proposed detection model significantly outperforms the version used-by MasaCHAI. This improvement is primarily attributed to the high-quality, task-specific dataset constructed in this work. Figure \ref{fig:confusion_matrix} presents the confusion matrix for our model, which highlights its strong performance in correctly classifying diverse circuit components.

\begin{table}[htbp]  
\centering  
\caption{Performance Comparison of Component Recognition Accuracy (\%)}  
\label{tab:component_accuracy_single}  
\small  

\begin{tabular}{lccccccccccccc}  
\toprule  
\textbf{Method}   & \textbf{AC\textsuperscript{1}}    & \textbf{BJT\textsuperscript{2}}   & \textbf{Batt\textsuperscript{3}}  & \textbf{Cap\textsuperscript{4}}   & \textbf{Cur\textsuperscript{5}} & \textbf{DC\textsuperscript{6}}    & \textbf{Diode\textsuperscript{7}} & \textbf{GND\textsuperscript{8}}   & \textbf{Ind\textsuperscript{9}}   & \textbf{MOS\textsuperscript{10}}   & \textbf{Res\textsuperscript{11}}   & \textbf{Volt\textsuperscript{12}} & \textbf{ALL\textsuperscript{13}}   \\ 
\midrule  
CHAI & 66.9          & 73.1          & \textbf{99.0 }         & 24.7          & 94.5         & 72.7          & \textbf{85.3}          & 1.7           & 51.4          & 63.2          & 80.1          & \textbf{70.2}          & 65.2          \\
Ours     & \textbf{79.0} & \textbf{82.4} & 85.3 & \textbf{95.7} & \textbf{99.4} & \textbf{92.9} & 83.1          & \textbf{99.1} & \textbf{87.0} & \textbf{99.4} & \textbf{95.4} & 60.3          & \textbf{88.3} \\ 
\bottomrule  
\end{tabular}
\par\vspace{2pt} 
\footnotesize
\textsuperscript{1} AC Source; \textsuperscript{2} Bipolar Junction Transistor; \textsuperscript{3} Battery; \textsuperscript{4} Capacitor; \textsuperscript{5} Current Source; \textsuperscript{6} DC Source; \textsuperscript{7} Diode; \textsuperscript{8} Ground; \textsuperscript{9} Inductor; \textsuperscript{10} MOSFET; \textsuperscript{11} Resistor; \textsuperscript{12} Voltage Source; \textsuperscript{13}All Classes.\\
\end{table}

\subsection{Ablation Experiment}

To systematically assess the contribution of individual components within the proposed framework, we conducted ablation experiments using Qwen-VL-Max as the core model. The corresponding results are summarized in Table \ref{tab:ablation}. In the table, "W/O CoT" refers to the removal of the CoT\ref{cot_method} mechanism; "W/O MICL" indicates that the proposed MICL strategy is replaced with standard single-modal In-Context Learning (ICL). This substitution is adopted instead of completely eliminating in-context learning, as removing contextual guidance altogether leads to highly inconsistent netlist formats that significantly complicate downstream parsing. "W/O Intent reasoning" refers to the removal of the intent reasoning\ref{intent reasoning} mechanism. The experimental results demonstrate that removing any core component results in a significant degradation  in the overall performance of the framework. 

However, while each component is indispensable, the extent of performance degradation upon its removal varies depending on circuit complexity. Removing the CoT mechanism reduces the average Pass@1 to 39.6\% and Pass@5 to 64.0\%. This mechanism is indispensable for complex circuits with intricate connectivity (Cases 7, 8, and 10), whereas it has minimal impact on simple standard circuits (Cases 1 and 2). This is because simple circuits do not require such a complex and rigorous reasoning process to successfully complete the image-to-netlist conversion. Replacing MICL with ICL reduces the average Pass@1 to 32.0\% and Pass@5 to 55.3\%. MICL is particularly effective in handling cases where image wiring is messy and complex, even if the underlying circuit logic is not. The advantage of this component is most evident in Case 7. Removing the Intent reasoning component leads to the most severe performance decline. This is because, without the error-correction mechanism, the framework's fault tolerance drops significantly. It becomes heavily reliant on the model generating a completely correct netlist in a single attempt and performing a correct circuit analysis to enable the Parameter Search Agent to accurately locate the appropriate parameter space, regardless of whether the cases are simple (Case 1,3) or complex (Case 9, 10, 11), this component greatly ensures the overall stability of the framework. Furthermore, for difficult cases (Cases 13–15), non-zero performance is observed only under the complete framework. This confirms that each design element plays a critical and complementary role in ensuring the effectiveness and robustness of the proposed framework.

\begin{table}[htbp]
\centering 
\caption{Experimental results of Pass@K metrics under different settings (simplified percentage)} 
\label{tab:ablation} 
\small 
\begin{tabular}{c|cc|cc|cc|cc}
\toprule 
\multirow{2}{*}{\textbf{Case}} & \multicolumn{2}{c|}{\textbf{W/O CoT}}               & \multicolumn{2}{c|}{\textbf{W/O MICL\textsuperscript{1}}}              & \multicolumn{2}{c|}{\textbf{W/O Intent reasoning}} & \multicolumn{2}{c}{\textbf{Full Framework}}        \\
                    & Pass@1 & Pass@5 & Pass@1 & Pass@5 & Pass@1 & Pass@5 & Pass@1 & Pass@5 \\ 
\midrule 
1                    & 86.7                       & \textbf{100.0}             & 80.0                       & \textbf{100.0}             & 33.3                       & 91.6                       & \textbf{100.0}             & \textbf{100.0}             \\
2                    & \textbf{100.0}             & \textbf{100.0}             & \textbf{100.0}             & \textbf{100.0}             & \textbf{100.0}             & \textbf{100.0}             & \textbf{100.0}             & \textbf{100.0}             \\
3                    & 40.0                       & 95.80                      & \textbf{46.7}              & \textbf{98.1}              & 6.7                        & 33.3                       & \textbf{46.7}              & \textbf{98.1}              \\
4                    & 80.0                       & 100.0                      & \textbf{93.3}              & \textbf{100.0}             & 86.7                       & \textbf{100.0}             & \textbf{93.3}              & \textbf{100.0}             \\
5                    & 80.0                       & \textbf{100.0}             & 33.3                       & 91.6                       & 46.7                       & 98.1                       & \textbf{100.0}             & \textbf{100.0}             \\
6                    & 86.7                       &\textbf{ 100.0  }                    & 60.0                       & 99.8                       & 80.0                       & \textbf{100.0}             & \textbf{93.3}              & \textbf{100.0}             \\
7                    & 6.7                        & 33.3                       & 0.0                        & 0.0                        & 6.7                        & 33.3                       & \textbf{20.0}              & \textbf{73.6}              \\
8                    & 6.7                        & 33.3                       & \textbf{13.3}              & \textbf{57.1}              & 0.0                        & 0.0                        & \textbf{13.3}              & \textbf{57.1}              \\
9                    & 33.3                       & 91.6                       & 13.3                       & 57.1                       & 6.7                        & 33.3                       & \textbf{60.0}              & \textbf{99.8}              \\

10                   & 6.7                        & 33.3                       & 0.0                        & 0.0                        & 26.7                       & 84.6                       & \textbf{33.3}              & \textbf{91.6}              \\
11                   & 46.7                       & 98.1                       & 33.3                       & 91.6                       & 33.3                       & 91.6                       & \textbf{93.3}              & \textbf{100.0}             \\
12                   & 20.0                       & 73.6                       & 6.7                        & 33.3                       & 13.3                       & 57.1                       & \textbf{33.3}              & \textbf{91.6}              \\
13                   & 0.0                        & 0.0                        & 0.0                        & 0.0                        & 0.0                        & 0.0                        & 0.0                        & 0.0                        \\
14                   & 0.0                        & 0.0                        & 0.0                        & 0.0                        & 0.0                        & 0.0                        & \textbf{6.7}                       & \textbf{33.3}                         \\
15                   & 0.0                        & 0.0                        & 0.0                        & 0.0                        & 0.0                        & 0.0                        & \textbf{6.7}                        & \textbf{33.3}                       \\ 
\midrule 
Avg                  & 39.6                       & 64.0                       & 32.0                       & 55.3                       & 29.3                       & 54.9                       & \textbf{53.3}              & \textbf{78.6}              \\
Solve                & 12                         & 12                         & 10                         & 10                         & 11                         & 11                         & \textbf{14}                & \textbf{14}                \\ 
\bottomrule 
\end{tabular}
\par\vspace{2pt} 
\footnotesize
\textsuperscript{1} Replace it with ICL;
\end{table}

\label{sec:experiment}

\section{Conclusion}
This paper presents \textbf{AnalogMaster}, a novel LLM-driven framework for end-to-end automation of analog IC design. The proposed framework first achieves accurate extraction of structurally complete netlists from circuit images through a joint reasoning mechanism. To enable downstream physical design, a parameter search agent is then introduced to transform isolated netlists into parameterized representations suitable for analog design workflows, overcoming the limitations of conventional sizing methods that rely on manually predefined parameter ranges. Finally, AnalogMaster integrates robust classical algorithms to sequentially perform device sizing optimization, automatic placement, and analog routing, forming a complete and coherent design pipeline. Experimental results demonstrate that AnalogMaster consistently delivers strong performance across multiple models. In particular, GPT-5 achieves success rates of 92.9\% and 99.9\% on the Pass@1 and Pass@5 metrics, respectively, validating the robustness and effectiveness of the proposed framework for full-flow analog IC design.  This work establishes a comprehensive paradigm for deeply integrating LLMs with traditional analog IC design algorithms and provides a practical foundation for highly reliable and fully automated analog IC design.


\label{sec:conclusion}

\section*{Funding}
This work was supported in part by the National Natural Science Foundation of China (92473207,92373102,62104174), State Key Laboratory of Integrated Chips and Systems (No.SKLICS-K202506), and National Science and Technology Major Project(No.2021ZD0114600).

\section*{CRediT authorship contribution statement}
Xianrong Qin: (Co-First Author), Investigation, Conceptualization, Methodology, Writing – original draft. Yong Zhang: (Co-First Author), Validation, Visualization. Ying Hu: Data collection, figure preparation. Tao Su: Visualization. Bo-Wen Jia: (Corresponding Author), Supervision, Project administration, Writing – review \& editing. Ning Xu:(Corresponding Author), Supervision, Project administration, Resources.
\section*{Declaration of competing interest}
The authors declare that they have no known competing financial interests or personal relationships that could have appeared to influence the work reported in this paper.

\section*{Data availability}
Data will be made available on request.

\section*{Acknowledgments}
The authors would like to thank Prof. Jun Hu, Prof. Keren Zhu for their valuable discussion.

\bibliographystyle{elsarticle-harv} 

\bibliography{reference_paper}

@article{JIN20253961,
title = {A Review of AI-Driven Automation Technologies: Latest Taxonomies, Existing Challenges, and Future Prospects},
journal = {Computers, Materials and Continua},
volume = {84},
number = {3},
pages = {3961-4018},
year = {2025},
issn = {1546-2218},
doi = {https://doi.org/10.32604/cmc.2025.067857},
url = {https://www.sciencedirect.com/science/article/pii/S1546221825007416},
author = {Weiqiang Jin and Ningwei Wang and Lei Zhang and Xingwu Tian and Bohang Shi and Biao Zhao},
}

@article{ROJEC201948,
title = {Analog circuit topology synthesis by means of evolutionary computation},
journal = {Engineering Applications of Artificial Intelligence},
volume = {80},
pages = {48-65},
year = {2019},
issn = {0952-1976},
doi = {https://doi.org/10.1016/j.engappai.2019.01.012},
url = {https://www.sciencedirect.com/science/article/pii/S0952197619300119},
author = {Žiga Rojec and Árpád Bűrmen and Iztok Fajfar},
}

@article{PAN2026113035,
title = {Graph Neural Networks Based Analog Circuit Link Prediction},
journal = {Engineering Applications of Artificial Intelligence},
volume = {163},
pages = {113035},
year = {2026},
issn = {0952-1976},
doi = {https://doi.org/10.1016/j.engappai.2025.113035},
url = {https://www.sciencedirect.com/science/article/pii/S0952197625030660},
author = {Guanyuan Pan and Tiansheng Zhou and Jianxiang Zhao and Zhi Li and Yugui Lin and Bingtao Ma and Yaqi Wang and Pietro Liò and Shuai Wang},
}

@article{chen2021evaluating,
  title={Evaluating large language models trained on code},
  author={Chen, Mark},
  journal={arXiv preprint arXiv:2107.03374},
  year={2021}
}

@article{xu2025paroute2,
  title={PARoute2: Enhanced Analog Routing via Performance-Drive Guidance Generation},
  author={Xu, Peng and Tu, Jindong and Chen, Guojin and Zhu, Keren and Chen, Tinghuan and Ho, Tsung-Yi and Yu, Bei},
  journal={IEEE Transactions on Computer-Aided Design of Integrated Circuits and Systems},
  year={2025},
  publisher={IEEE}
}

@inproceedings{xu2017hierarchical,
  title={Hierarchical and analytical placement techniques for high-performance analog circuits},
  author={Xu, Biying and Li, Shaolan and Xu, Xiaoqing and Sun, Nan and Pan, David Z},
  booktitle={Proceedings of the 2017 ACM on International Symposium on Physical Design},
  pages={55--62},
  year={2017}
}

@article{ma2010simultaneous,
  title={Simultaneous handling of symmetry, common centroid, and general placement constraints},
  author={Ma, Qiang and Xiao, Linfu and Tam, Yiu-Cheong and Young, Evangeline FY},
  journal={IEEE Transactions on Computer-Aided Design of Integrated Circuits and Systems},
  volume={30},
  number={1},
  pages={85--95},
  year={2010},
  publisher={IEEE}
}

@inproceedings{lyu2018batch,
  title={Batch Bayesian optimization via multi-objective acquisition ensemble for automated analog circuit design},
  author={Lyu, Wenlong and Yang, Fan and Yan, Changhao and Zhou, Dian and Zeng, Xuan},
  booktitle={International conference on machine learning},
  pages={3306--3314},
  year={2018},
  organization={PMLR}
}

@inproceedings{somayaji2021prioritized,
  title={Prioritized reinforcement learning for analog circuit optimization with design knowledge},
  author={Somayaji, NS Karthik and Hu, Hanbin and Li, Peng},
  booktitle={2021 58th ACM/IEEE Design Automation Conference (DAC)},
  pages={1231--1236},
  year={2021},
  organization={IEEE}
}

@article{zhang2025reinforcement,
  title={Reinforcement Learning-Driven Net Order Selection for Efficient Analog IC Routing},
  author={Zhang, Yong and Li, Wen-Jie and Ge, Guo-Jing and Wang, Jin-Qiao and Jia, Bo-Wen and Xu, Ning},
  journal={Integration},
  pages={102623},
  year={2025},
  publisher={Elsevier}
}

@article{zhang2025pin,
  title={Pin Access Planning-driven Matching-based Analog Integrated Circuits Routing},
  author={Zhang, Yong and Liu, Ao and Yin, Yong and Xu, Ning and Jia, Bowen},
  journal={Microelectronics Journal},
  pages={106960},
  year={2025},
  publisher={Elsevier}
}

@inproceedings{min2022metaicl,
  title={Metaicl: Learning to learn in context},
  author={Min, Sewon and Lewis, Mike and Zettlemoyer, Luke and Hajishirzi, Hannaneh},
  booktitle={Proceedings of the 2022 conference of the North American chapter of the Association for Computational Linguistics: Human Language Technologies},
  pages={2791--2809},
  year={2022}
}

@article{liu2025layoutcopilot,
  title={Layoutcopilot: An llm-powered multi-agent collaborative framework for interactive analog layout design},
  author={Liu, Bingyang and Zhang, Haoyi and Gao, Xiaohan and Kong, Zichen and Tang, Xiyuan and Lin, Yibo and Wang, Runsheng and Huang, Ru},
  journal={IEEE Transactions on Computer-Aided Design of Integrated Circuits and Systems},
  year={2025},
  publisher={IEEE}
}

@article{liu2024ladac,
  title={Ladac: Large language model-driven auto-designer for analog circuits},
  author={Liu, Chengjie and Liu, Yijiang and Du, Yuan and Du, Li},
  journal={Authorea Preprints},
  year={2024},
  publisher={Authorea}
}

@inproceedings{zhang2025analogxpert,
  title={Analogxpert: Automating analog topology synthesis by incorporating circuit design expertise into large language models},
  author={Zhang, Haoyi and Sun, Shizhao and Lin, Yibo and Wang, Runsheng and Bian, Jiang},
  booktitle={2025 International Symposium of Electronics Design Automation (ISEDA)},
  pages={772--777},
  year={2025},
  organization={IEEE}
}

@article{somayaji2025llm-uso,
  title={LLM-USO: Large Language Model-based Universal Sizing Optimizer},
  author={Somayaji, NS Karthik and Li, Peng},
  journal={IEEE Transactions on Computer-Aided Design of Integrated Circuits and Systems},
  year={2025},
  publisher={IEEE}
}

@inproceedings{vungarala2024spicepilot,
  title={Spicepilot: Navigating spice code generation and simulation with ai guidance},
  author={Vungarala, Deepak and Alam, Sakila and Ghosh, Arnob and Angizi, Shaahin},
  booktitle={2024 IEEE International Conference on Rebooting Computing (ICRC)},
  pages={1--6},
  year={2024},
  organization={IEEE}
}

@article{liu2023chipnemo,
  title={Chipnemo: Domain-adapted llms for chip design},
  author={Liu, Mingjie and Ene, Teodor-Dumitru and Kirby, Robert and Cheng, Chris and Pinckney, Nathaniel and Liang, Rongjian and Alben, Jonah and Anand, Himyanshu and Banerjee, Sanmitra and Bayraktaroglu, Ismet and others},
  journal={arXiv preprint arXiv:2311.00176},
  year={2023}
}

@article{bhandari2024llm-aided-testbench,
  title={Llm-aided testbench generation and bug detection for finite-state machines},
  author={Bhandari, Jitendra and Knechtel, Johann and Narayanaswamy, Ramesh and Garg, Siddharth and Karri, Ramesh},
  journal={arXiv preprint arXiv:2406.17132},
  year={2024}
}

@inproceedings{qiu2024autobench,
  title={Autobench: Automatic testbench generation and evaluation using llms for hdl design},
  author={Qiu, Ruidi and Zhang, Grace Li and Drechsler, Rolf and Schlichtmann, Ulf and Li, Bing},
  booktitle={Proceedings of the 2024 ACM/IEEE International Symposium on Machine Learning for CAD},
  pages={1--10},
  year={2024}
}

@article{thakur2024verigen,
  title={Verigen: A large language model for verilog code generation},
  author={Thakur, Shailja and Ahmad, Baleegh and Pearce, Hammond and Tan, Benjamin and Dolan-Gavitt, Brendan and Karri, Ramesh and Garg, Siddharth},
  journal={ACM Transactions on Design Automation of Electronic Systems},
  volume={29},
  number={3},
  pages={1--31},
  year={2024},
  publisher={ACM New York, NY}
}

@inproceedings{lu2024rtllm,
  title={Rtllm: An open-source benchmark for design rtl generation with large language model},
  author={Lu, Yao and Liu, Shang and Zhang, Qijun and Xie, Zhiyao},
  booktitle={2024 29th Asia and South Pacific Design Automation Conference (ASP-DAC)},
  pages={722--727},
  year={2024},
  organization={IEEE}
}

@article{thakur2023autochip,
  title={Autochip: Automating hdl generation using llm feedback},
  author={Thakur, Shailja and Blocklove, Jason and Pearce, Hammond and Tan, Benjamin and Garg, Siddharth and Karri, Ramesh},
  journal={arXiv preprint arXiv:2311.04887},
  year={2023}
}

@inproceedings{liu2023verilogeval,
  title={Verilogeval: Evaluating large language models for verilog code generation},
  author={Liu, Mingjie and Pinckney, Nathaniel and Khailany, Brucek and Ren, Haoxing},
  booktitle={2023 IEEE/ACM International Conference on Computer Aided Design (ICCAD)},
  pages={1--8},
  year={2023},
  organization={IEEE}
}

@article{wu2024chateda,
  title={Chateda: A large language model powered autonomous agent for eda},
  author={Wu, Haoyuan and He, Zhuolun and Zhang, Xinyun and Yao, Xufeng and Zheng, Su and Zheng, Haisheng and Yu, Bei},
  journal={IEEE Transactions on Computer-Aided Design of Integrated Circuits and Systems},
  volume={43},
  number={10},
  pages={3184--3197},
  year={2024},
  publisher={IEEE}
}

@article{liu2024rtlcoder,
  title={Rtlcoder: Fully open-source and efficient llm-assisted rtl code generation technique},
  author={Liu, Shang and Fang, Wenji and Lu, Yao and Wang, Jing and Zhang, Qijun and Zhang, Hongce and Xie, Zhiyao},
  journal={IEEE Transactions on Computer-Aided Design of Integrated Circuits and Systems},
  year={2024},
  publisher={IEEE}
}

@article{chang2024lamagic,
  title={Lamagic: Language-model-based topology generation for analog integrated circuits},
  author={Chang, Chen-Chia and Shen, Yikang and Fan, Shaoze and Li, Jing and Zhang, Shun and Cao, Ningyuan and Chen, Yiran and Zhang, Xin},
  journal={arXiv preprint arXiv:2407.18269},
  year={2024}
}

@article{vaswani2017attention,
  title={Attention is all you need},
  author={Vaswani, Ashish and Shazeer, Noam and Parmar, Niki and Uszkoreit, Jakob and Jones, Llion and Gomez, Aidan N and Kaiser, {\L}ukasz and Polosukhin, Illia},
  journal={Advances in neural information processing systems},
  volume={30},
  year={2017}
}

@article{liu2024ampagent,
  title={Ampagent: An llm-based multi-agent system for multi-stage amplifier schematic design from literature for process and performance porting},
  author={Liu, Chengjie and Chen, Weiyu and Peng, Anlan and Du, Yuan and Du, Li and Yang, Jun},
  journal={arXiv preprint arXiv:2409.14739},
  year={2024}
}

@article{gao2025analoggenie,
  title={AnalogGenie: A generative engine for automatic discovery of analog circuit topologies},
  author={Gao, Jian and Cao, Weidong and Yang, Junyi and Zhang, Xuan},
  journal={arXiv preprint arXiv:2503.00205},
  year={2025}
}

@article{masalaCHAI-2025,
  title={Masala-chai: A large-scale spice netlist dataset for analog circuits by harnessing ai},
  author={Bhandari, Jitendra and Bhat, Vineet and He, Yuheng and Rahmani, Hamed and Garg, Siddharth and Karri, Ramesh},
  journal={arXiv preprint arXiv:2411.14299},
  year={2024}
}

@inproceedings{lai2025analogcoder,
  title={Analogcoder: Analog circuit design via training-free code generation},
  author={Lai, Yao and Lee, Sungyoung and Chen, Guojin and Poddar, Souradip and Hu, Mengkang and Pan, David Z and Luo, Ping},
  booktitle={Proceedings of the AAAI Conference on Artificial Intelligence},
  volume={39},
  pages={379--387},
  year={2025}
}

@Misc{prokComputerVisionDataset,
	howpublished = {\url{https://universe.roboflow.com/anotnio/prok}},
	note = {[Available:2025]},
	title = {prok Computer Vision Dataset},
	author = {anotnio},
    year = 2024,
	}

@Misc{MOSFETDetectionComputerVisionDataset,
	howpublished = {\url{https://universe.roboflow.com/un-eiyzu/mosfet-detection-nrqmg}},
	note = {[Available:2025]},
	title = {MOSFET Detection Computer Vision Dataset},
	author = {Un},
    year = 2025,
	}

@Misc{CNN_CircuitComputerVisionModel,
	howpublished = {\url{https://universe.roboflow.com/circuit-circuit-do-circuit/cnn_circuit
}},
	note = {[Available:2025]},
	title = {CNN Circuit Computer Vision Model},
	author = {{Circuit Circuit Do Circuit}},
    year = 2022,
	}

@Misc{v3qwe,
	howpublished = {\url{https://universe.roboflow.com/project-gtqqq/v3qwe}},
	note = {[Available:2025]},
	title = {v3qwe Computer Vision Dataset},
	author = {project-gtqqq},
    year = 2022,
	}

\end{document}